\documentclass[11pt,a4paper,draft]{article}
\usepackage{amssymb}
\usepackage[final]{graphicx}

\usepackage{newpxtext}
\usepackage[varbb]{newpxmath}
\usepackage{sourcesanspro}

\usepackage{sectsty}
\allsectionsfont{\sffamily}

\usepackage[margin=1in]{geometry}

\usepackage[activate={true,nocompatibility}]{microtype}

\usepackage{enumitem}

\usepackage{setspace}

\usepackage{hyperref}
\hypersetup{bookmarksopen=true,bookmarksnumbered=true,hidelinks,final=true}

\usepackage{doi}

\usepackage[style=authoryear-comp, 
            backend=biber,
						natbib=true,
            urldate=comp,
            maxcitenames=2,
            mincitenames=1,
            maxbibnames=3,
						uniquename=false,
            giveninits=true]{biblatex}
\DeclareLabeldate{%
  \field{date}
  \field{pubstate}
}

\usepackage{bm}

\DeclareMathOperator*{\cor}{cor}

\newcommand{\dd}{\mathop{}\!d}

\DeclareMathOperator{\Normal}{N}
\DeclareMathOperator{\halfNormal}{half-N}
\DeclareMathOperator{\MVN}{MVN}

\newcommand*{\tr}{^{\scriptscriptstyle\mathsf{T}}}

\usepackage{mleftright}

\usepackage{pdflscape}

\usepackage[capitalise,noabbrev]{cleveref}

\usepackage{subfig}

\usepackage{array,booktabs}

\usepackage[dvipsnames]{xcolor}
\usepackage{todonotes}

\colorlet{dpcolor}{NavyBlue!25}
\colorlet{sdcolor}{ForestGreen!25}
\colorlet{nwcolor}{Maroon!25}
\colorlet{tacolor}{BurntOrange!25}

\usepackage{soul}

\newcounter{dpnote}

\newcounter{sdnote}

\newcounter{nwnote}

\newcounter{tanote}

\begin{document}
\singlespacing
\thispagestyle{empty}
\begin{titlepage}
	\begin{center}\LARGE\sffamily\bfseries
		Network Meta-Analysis of survival outcomes with non-proportional hazards using flexible M-splines
	\end{center}
	\vspace{2\baselineskip}
	\begin{raggedright}\footnotesize
		David M.\ Phillippo\footnote{University of Bristol, Canynge Hall, 39 Whatley Road, Bristol, BS8 2PS, UK. Email:	david.phillippo@bristol.ac.uk}\\
		University of Bristol, UK\\[1em]
		Ayman Sadek\\
		University of Bristol, UK\\[1em]
		Hugo Pedder\\
		University of Bristol, UK\\[1em]
		Nicky J.\ Welton\\
		University of Bristol, UK
	\end{raggedright}
	\vspace{2\baselineskip}
  \subsection*{Abstract}
	Network meta-analysis (NMA) is widely used in healthcare decision-making, where estimates of the effect of multiple treatments on outcomes are required.
	For time-to-event outcomes such as survival or disease progression the most common approach is to model log hazard ratios; however, this relies on the proportional hazards assumption.
	Novel treatments such as immunotherapies are expected to display complex hazard functions that cannot be captured by standard parametric models, which results in non-proportional hazards when comparing treatments from different classes.
	As a result, alternative models such as fractional polynomials or restricted cubic splines are often used.
	These  allow substantial flexibility on the shape of the baseline hazard, but require time-consuming model selection or are intractable for Bayesian analysis.

	We propose a flexible NMA model using M-splines on the baseline hazard, with a novel weighted random walk prior distribution that provides shrinkage to avoid overfitting and is invariant to the choice of knots and timescale.
	Non-proportional hazards are modelled either by stratifying by treatment or  by introducing treatment effects on the spline coefficients, and covariates may be included on the log hazard rate and spline coefficients.
	Treatment and covariate effects on the spline coefficients are given random walk prior distributions to smoothly model departures from proportionality over time.
	The methods are implemented in the user-friendly R package multinma, which supports analyses with aggregate data, individual participant data, or mixtures of both.
	We apply the methods to a NMA of progression-free survival with treatments for non-small cell lung cancer. 

	\paragraph*{Keywords:} network meta-analysis, indirect comparison, survival analysis, proportional hazards, time-to-event data
\end{titlepage}

\clearpage
\doublespacing
\section{Introduction}\label{sec:introduction}

Survival or time-to-event outcomes are commonplace across healthcare decision-making, not least in fields such as oncology.
Often, decision-makers are interested in absolute survival estimates in a target population, for example as inputs to an economic model.
Flexible parametric models are widely used for this purpose since they place no restriction on the form of the baseline hazard, unlike parametric models that are restricted to simpler hazard trajectories.
This is particularly important where hazard functions are expected to be complex such as with cancer immunotherapies \parencite{TSD21,Palmer2023}.
A range of flexible methods have been proposed, including restricted cubic splines \parencite[the \emph{Royston-Parmar} model;][]{Royston2002}, M-splines \parencite{Brilleman2020}, fractional polynomials \parencite{Royston1994}, and piecewise exponential via Poisson Generalised Linear Models \parencite{Lambert2005}.

When multiple randomised controlled trials are available on treatments of interest, network meta-analysis (NMA) is used to coherently combine evidence on relative effectiveness from all studies \parencite{TSD2, Higgins1996, Lu2004, Dias2018}.
The simplest approach to NMA of survival outcomes is a contrast-based analysis combining summary log hazard ratios from each study with a Normal likelihood.
However, this approach cannot provide absolute predictions unless external information on a suitable baseline hazard function is available, and it does not allow the proportional hazards assumption to be relaxed.
More sophisticated approaches to synthesis of survival outcomes within the NMA framework have been proposed, such as parametric NMA models \parencite{Ouwens2010}, fractional polynomial NMA models \parencite{Jansen2011}, and Royston-Parmar NMA models \parencite{Freeman2017}.
These require individual event and censoring times to be available from each of the studies in the analysis.
When the original individual participant data are unavailable, these may be reconstructed by digitising published Kaplan-Meier curves followed by the algorithm of \textcite{Guyot2012}.

Parametric NMA models \parencite{Ouwens2010} jointly synthesise shape and rate parameters from parametric survival curves estimated within each study (e.g.\ with Weibull, Gompertz, or log-Normal distributions).
Non-proportional hazards may be modelled by including a treatment effect on the shape parameter, as well as the treatment effect on the log hazard rate.
However, the form of the baseline hazard is still restricted to follow a parametric distribution, which may not always be suitable.

The most widely-used flexible approach is the fractional polynomial NMA model proposed by \textcite{Jansen2011}.
A wide range of shapes for the baseline hazard function are permitted by including coefficients for different powers of time in a linear predictor on the log hazard.
When treatment effects are placed on these coefficients a non-proportional hazards model is obtained, as the log hazard on each treatment is allowed to follow a different polynomial trajectory.
The primary disadvantage of this approach lies in the necessary but time-consuming selection of suitable powers, since the model must be re-run with each choice of powers and the best-fitting model chosen.
With a first-order fractional polynomial there are eight models to select from (with the usual set of powers $-2$, $-1$, $-0.5$, $0$, $0.5$, $1$, $2$, $3$); with a second-order fractional polynomial there are a further 36 models.
The fractional polynomial NMA model has no closed-form likelihood, but is typically fit in a Bayesian framework using an approximate piecewise binomial likelihood.

The Royston-Parmar cubic spline model \parencite{Royston2002} is widely-used for survival modelling and has been proposed for a NMA setting by \textcite{Freeman2017}.
A restricted cubic spline is placed on the log cumulative hazard, allowing this to vary smoothly over time.
\textcite{Freeman2017} implement the model in a Bayesian framework with independent vague prior distributions on the spline coefficients.
Non-proportional hazards can be modelled by including an interaction between treatment and $\log(\textrm{time})$ in the linear predictor.
However, this model is often very difficult to fit in practice because the restricted cubic splines do not constrain the cumulative hazard to be monotonically increasing \parencite[a property noted by][]{Royston2002}.
In a frequentist setting this is typically of little consequence, since the data are usually sufficient to ensure that the maximum likelihood estimate is in a plausible region of the parameter space.
However, in a Bayesian setting the entire parameter space must be considered---including the implausible regions---making sampling and estimation problematic.

M-splines, proposed for the analysis of survival outcomes outside the NMA setting by \textcite{Brilleman2020}, offer an attractive alternative to these previous approaches.
These are placed directly on the baseline hazard, which ensures that the cumulative hazards are strictly increasing and have a known closed form as integrated M-splines (known as \emph{I-splines}).
\citeauthor{Brilleman2020} placed a Dirichlet prior distribution on the spline coefficients.
However, this has no shrinkage properties which means that model fit is undesirably sensitive to the number and location of knots.
\textcite{Jackson2023} proposed a logistic random effect prior distribution for the spline coefficients, in an attempt to provide shrinkage and avoid overfitting.
In practice however, we have found that the level of shrinkage is not sufficient and that increasing the number of knots still leads to overfitting.
Previously, we applied M-splines in a multilevel network meta-regression setting (a generalisation of NMA to coherently incorporate both individual- and aggregate-level data), using a novel weighted random walk prior distribution for the spline coefficients that does lead to appropriate levels of shrinkage \parencite{Phillippo2024_survival}.
This approach is yet to be described in detail and requires extension to model non-proportional hazards.

In this paper, we propose a flexible NMA model using M-splines on the baseline hazard, and extend this to allow non-proportional hazards to be modelled.
We begin by outlining a motivating case study of progression free survival (PFS) on treatments for non-small cell lung cancer (NSCLC).
We then describe the M-spline NMA model in detail, before applying this to the NSCLC example.
The proposed models are implemented in the \emph{multinma} R package \citep{multinma}; full analysis code and data are available from the GitHub repository \href{https://github.com/dmphillippo/mspline-survival-paper}{dmphillippo/mspline-survival-paper}.
We conclude with a discussion including comparisons with alternative approaches.

\section{Case study: progression free survival with non-small cell lung cancer}\label{sec:case_study}

\begin{figure}
	\centering
  \includegraphics[width=0.8\textwidth]{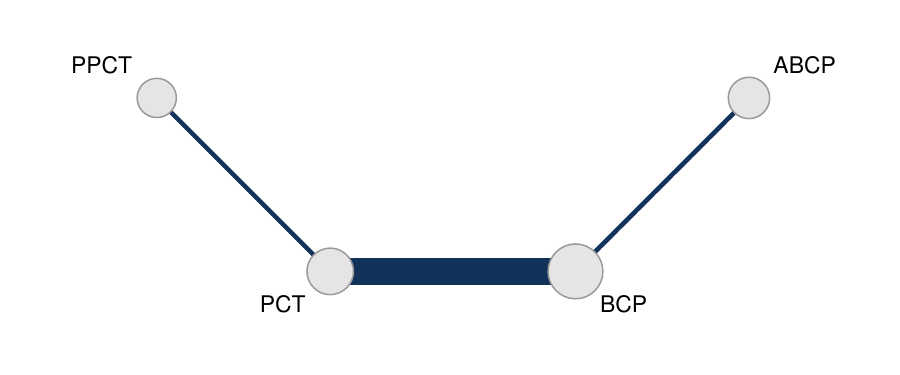}
  \caption{Network of four studies comparing first-line treatments for non-small cell lung cancer. Edge widths and numbers indicate the number of studies making each comparison, and the size of each node corresponds to the number of individuals randomised to each treatment.}
  \label{fig:nsclc_network}
\end{figure}

\begin{figure}
	\centering
  \includegraphics[width=0.8\textwidth]{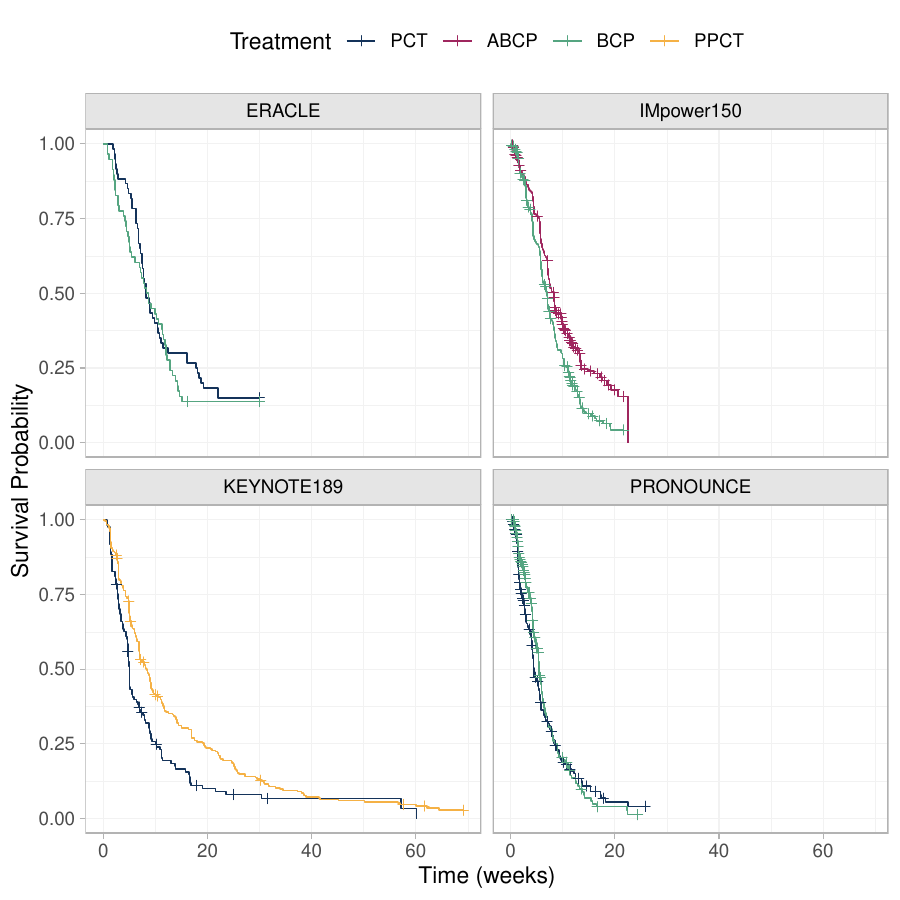}
  \caption{Kaplan-Meier curves of progression free survival on each treatment in each trial.}
  \label{fig:nsclc_km}
\end{figure}

\Cref{fig:nsclc_network} shows a network of four studies comparing first-line treatments for advanced/metastatic non-small cell lung cancer, for patients with non-squamous cell carcinomas with programmed death-ligand 1 (PDL-1) expression less than 50\%.
Three non-targeted therapies were considered for a treatment decision: pembrolizumab with pemetrexed and platinum chemotherapy (PPCT); atezolizumab plus bevacizumab, carboplatin and paclitaxel (ABCP); and pemetrexed with platinum doublet chemotherapy (PCT).
A fourth, bevacizumab with carboplatin and paclitaxel (BCP), was included as a comparator in the NMA to connect the network.
Kaplan-Meier curves of progression free survival from each study were digitised and the algorithm of \textcite{Guyot2012} used to obtain reconstructed event/censoring times, which are shown in \cref{fig:nsclc_km}.
These studies were identified and data extracted as part of a pilot project for the National Institute for Health and Care Excellence \parencite{Pedder2025}.

\begin{figure}
	\centering
  \includegraphics[width=0.8\textwidth]{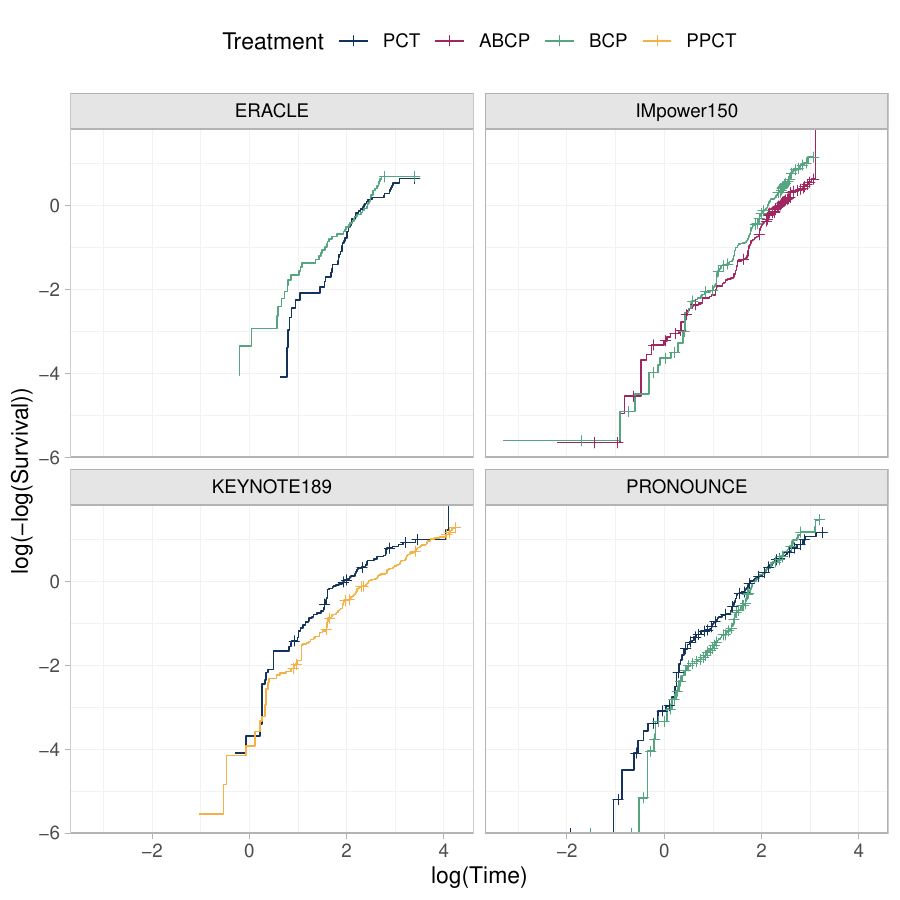}
  \caption{Complementary log-log plot of the Kaplan-Meier survival estimate against $\log(\textrm{time})$. Non-parallel curves indicate that hazards are non-proportional.}
  \label{fig:nsclc_loglog}
\end{figure}

The mode of action of immunotherapies like pembrolizumab and atezolizumab means that treatment effects are typically delayed whilst the immune system activates and proliferates a response.
This motivates the use of a flexible baseline hazard model to capture the complex behaviour of the baseline hazard.
Moreover, as a result we also expect to find evidence of non-proportional hazards in the included studies, as the shape of the baseline hazard is expected to differ between treatment arms with different modes of action.
Examining the log-log plots in \cref{fig:nsclc_loglog}, we indeed see that the curves are not parallel and thus display non-proportional hazards.
Lastly, all treatments of interest are not compared in a single study, but instead form a connected network of evidence (\cref{fig:nsclc_network}).
We therefore require an approach that allows for modelling of flexible baseline hazard functions and non-proportional hazards within the NMA framework.

\section{Methods}\label{sec:methods}

Consider a network of $J$ randomised controlled trials, each investigating a subset $\mathscr{K}_j$ of $K$ treatments.
Each study provides pairs $y_{ijk} = \lbrace t_{ijk}, c_{ijk} \rbrace$ of outcome times $t_{ijk}$ and censoring indicators $c_{ijk}$ for each individual $i$ in study $j$ receiving treatment $k$, where $c_{ijk}=1$ if an individual experiences the event or $c_{ijk}=0$ if they are censored.
Where such data are not available first-hand (as is often the case in practice), these may be obtained by digitizing published Kaplan-Meier curves and reconstructing the event and censoring times using an algorithm such as that described by \textcite{Guyot2012}.

A general proportional hazards network meta-analysis model for survival data takes the form
\begin{subequations}
\begin{align}
	h_{jk}(t) &= h_{0,j}(t) \exp(\eta_{jk}) \\
	\eta_{jk} &= \mu_j + \delta_{jk}
\end{align}
\end{subequations}
where $h_{jk}(t)$ is the hazard function at time $t$ on treatment $k$ in study $j$ and $h_{0,j}(t)$ is a baseline hazard function.
The linear predictor $\eta_{jk}$ on the log hazard scale includes a study-specific intercept $\mu_j$, and a study-specific relative treatment effect $\delta_{jk}$ for treatment $k$ against the network reference treatment 1.
For a fixed effect model we have $\delta_{jk} = d_k$, where $d_k$ is the relative treatment effect for treatment $k$ vs.\ treatment 1.
For a random effects model we have $\delta_{jk} \sim \Normal(d_k, \tau^2)$, where $\tau$ is the between-studies standard deviation, assumed common between comparisons.
Random effects for studies with two or more non-reference treatment arms are correlated and modelled using a multivariate Normal distribution with $\cor(\delta_{ja}, \delta_{jb}) = 0.5$ for all $a,b > 1$ under the assumption of common heterogeneity variance \parencite{Higgins1996}.
We set $\delta_{j1}=0$ and $d_1=0$.

In this network meta-analysis parameterisation, the consistency equations $d_{ab} = d_b - d_a$ describing relative effects between any two treatments are implicit and arise from additivity on the linear predictor scale.
Inconsistency can be assessed using unrelated mean effects or node-splitting models \citep{TSD4}.
An unrelated mean effects model replaces $\delta_{jk}$ with $\delta_{a_{j1}k}$, where $a_{j1}$ is the treatment in arm 1 of study $j$, which is then either equal to $d_{a_{j1}k}$ in a fixed effects model or distributed as $\Normal(d_{a_{j1}k}, \tau^2)$ in a random effects model.
Each relative effect $d_{ab}$ is given an independent prior distribution; the consistency equations are not invoked.
A node-splitting model for the $b$ vs.\ $a$ comparison replaces the linear predictor $\eta_{jk}$ for any study with both $a$ and $b$ arms by $\eta_{jk} = \mu_j + \delta_{jk} + \mathbb{I}(k = b) \vartheta_{ab}$.
The inconsistency factor $\vartheta_{ab} = d^\mathrm{dir}_{ab} - d_{ab}$ is the difference between the direct and indirect evidence on this comparison.

The likelihood for each event or censoring time is given by
\begin{equation}
	L_{ijk} = S_{jk}(t_{ijk}) h_{jk}(t_{ijk})^{c_{ijk}},
\end{equation}
where $S_{jk}(t)$ is the survival function,
\begin{equation}
	S_{jk}(t) = \exp \mleft( - \int_0^t h_{jk}(u) \mright) \dd u.
\end{equation}

Survival and hazard functions for a range of parametric survival models are described elsewhere by several authors, including \textcite{Brilleman2020} and \textcite{Phillippo2024_survival}.

\subsection{M-splines for flexible baseline hazards}

We propose a model using flexible M-splines for the baseline hazard, with survival and hazard functions
\begin{subequations}\label{eqn:mspline_ph}
\begin{align}
	S_{jk}(t) &= \exp\mleft( -\bm{\alpha}_j\tr \bm{I}_\kappa(t, \bm{\zeta}_j) \exp(\eta_{jk}) \mright) \label{eqn:mspline_ph_S}\\
	h_{jk}(t) &= \bm{\alpha}_j\tr \bm{M}_\kappa(t, \bm{\zeta}_j) \exp(\eta_{jk}) \label{eqn:mspline_ph_h}
\end{align}
\end{subequations}
where $\bm{\alpha}_j$ is a study-specific vector of spline coefficients, $\bm{M}_\kappa(t, \bm{\zeta}_j)$ is the M-spline basis of order $\kappa$ with a study-specific knot sequence $\bm{\zeta}_j$ evaluated at time $t$, and $\bm{I}_\kappa(t, \bm{\zeta}_j)$ is the corresponding integrated M-spline basis (an I-spline basis).
The basis polynomials have degree $\kappa-1$, so a basis of order 4 corresponds to a cubic M-spline basis.
Notice that we stratify the baseline hazard by study to respect randomisation, akin to the stratification of the study-specific intercepts $\mu_j$ in the linear predictor.

The knot sequence $\bm{\zeta}_j = (\zeta_{j,0}, \cdots, \zeta_{j, L+1})$ is a strictly increasing vector of length $L+2$, where $L$ is the number of internal knots chosen by the analyst, and $\zeta_{j,0}$ and $\zeta_{j,L+1}$ are the lower and upper boundary knots.
Internal knots may be placed at arbitrary locations within the boundary knots.
By default, we choose to place these at evenly-spaced quantiles of the observed event times in each study, and place boundary knots at time 0 and the last event/censoring time in each study.
The dimension of the spline basis, i.e. the number of spline coefficients in the vector $\bm{\alpha}_j$, is equal to $L+\kappa$.

The M-spline and I-spline bases are constructed using the recursive formulae of \citet{Ramsay1988}, detailed in \cref{sec:mspline_basis}, which are implemented in the \textit{splines2} R package \citep{splines2}.

\subsection{Random walk shrinkage prior}\label{sec:methods_rw_prior}

The spline coefficients $\bm{\alpha}_j$ lie in the unit simplex, i.e.\ $0 \le \alpha_{j,s} \le 1$ $\forall s = 1,\dots,L+\kappa$, and $\sum_{s=1}^{L+\kappa} \alpha_{j,s}=1$.
To avoid overfitting, we propose the use of a novel weighted random walk prior distribution on the inverse-softmax transformed spline coefficients.
In practice, this means that the analyst simply needs to specify a sufficiently-large number of knots (we choose 7 internal knots as the default in \textit{multinma}), which are then smoothed over time and shrunk towards a constant baseline hazard.
The amount of smoothing and shrinkage is controlled by the standard deviation of the random walk, which is given a prior distribution and estimated from the data.

This weighted random walk prior distribution is defined as follows:
\begin{subequations}\label{eqn:rw1_prior}
\begin{align}
	\bm{\alpha}_j &= \operatorname{softmax}(\bm{\alpha_j^*}) \\
	\alpha^*_{j,l} &= \varphi_{j,l} + \sum_{m = 1}^l u_{j,m} \quad \forall l=1,\dots,L+\kappa-1\\
	u_{j,l} &\sim \Normal(0, \sigma^2_j w_{j,l}) \quad \forall l=1,\dots,L+\kappa-1
\end{align}
\end{subequations}
where the softmax function maps a real-valued vector $\bm{\alpha_j^*}$ of length $L+\kappa-1$ to the coefficient vector $\bm{\alpha}_j$ of length $L+\kappa$ on the unit simplex:
\begin{equation}
	\operatorname{softmax}(\bm{\alpha}^*_j) = \frac{\mleft\lbrack 1, \exp(\bm{\alpha}^*_j)\tr \mright\rbrack\tr}{1 + \sum_{l=1}^{L+\kappa-1} \exp(\alpha^*_{j,l})}
\end{equation}
with inverse
\begin{equation}
	\operatorname{softmax}^{-1}(\bm{\alpha}_j) = \log(\bm{\alpha}_{j,2:(L+\kappa)}) - \log(\alpha_{j,1}).
\end{equation}
Rather than having zero mean, this random walk is centred around a prior mean vector $\bm{\varphi}_j=(\varphi_{j,1},\dots,\varphi_{j,L+\kappa-1})$ which corresponds to a constant baseline hazard:
\begin{equation}
	\bm{\varphi}_j = \operatorname{softmax}^{-1}\mleft( \frac{\bm{\zeta}^*_{j,\kappa:(L+2\kappa)} - \bm{\zeta}^*_{j,1:(L+\kappa)}}{\kappa (\zeta_{j,L+1} - \zeta_{j,0})} \mright)
\end{equation}
where $\bm{\zeta}^*_j$ is the augmented knot vector (see \cref{sec:mspline_basis}).
This idea borrows from \citet{Jackson2023}, who used $\bm{\varphi}_j$ as the mean for a random effect prior on $\bm{\alpha}_j$.
The random walk standard deviation $\sigma_j$ controls the amount of smoothing and shrinkage; as $\sigma_j$ approaches zero the baseline hazard becomes smoother (less ``wiggly'') and approaches a constant baseline hazard.
Furthermore, we introduce weights $\bm{w}_j$ into the random walk which are defined by the distance between each pair of knots:
\begin{equation}\label{eqn:mspline_rw1_weights}
	\bm{w}_j = \frac{\bm{\zeta}^*_{j,(\kappa+1):(L+2\kappa-1)} - \bm{\zeta}^*_{j,2:(L+\kappa)}}{(\kappa-1)(\zeta_{j,L+1} - \zeta_{j,0})}.
\end{equation}
This follows a similar approach to the Bayesian P-splines proposed by \citet{Li2022}, except that here we additionally normalise the weights to sum to 1.

These weights serve two purposes.
Firstly, the weights allow the overall prior distribution on the baseline hazard to be invariant to the knot locations $\bm{\zeta}_j$, even when these are unevenly spaced over time.
Without the weights, time periods with a greater number of knots (typically the start of follow-up, when we choose the knot locations by quantiles of observed survival times) will have greater prior variation, resulting in under-smoothing \citep{Li2022}.
Secondly, normalising the weights controls the overall variation in the prior, making the overall prior distribution on the baseline hazard invariant to the number of knots and to the timescale (i.e.\ $\zeta_{j,L+1} - \zeta_{j,0}$).
Without normalising, the overall variation in the prior increases as the number of knots increases or as the timescale decreases; increasing the number of knots or decreasing the timescale (e.g.\ counting time in weeks instead of days) would require a tighter prior distribution on $\sigma_j$ to achieve the same level of smoothing. 
These weights greatly simplify specification of a prior distribution for $\sigma_j$, as its interpretation no longer depends on the number or spacing of the knots, or on the timescale.
We choose $\sigma_j \sim \halfNormal(0, 1^2)$ as a weakly-informative prior distribution by default.
This results in prior baseline hazard trajectories where 95\% vary within approximately a factor of 12 between their highest and lowest values.

Degree-zero M-splines ($\kappa = 1$) correspond to a piecewise exponential baseline hazard.
However, the weights $\bm{w}_j$ are degenerate when $\kappa=1$, as equation \eqref{eqn:mspline_rw1_weights} then results in zero divided by zero.
In \cref{sec:rw_prior_pexp}, we propose a modification of the weights for the piecewise exponential case that retains the attractive shrinkage properties and is approximately invariant to the number and location of knots.

\begin{figure}
	\centering
  \includegraphics[width=\textwidth]{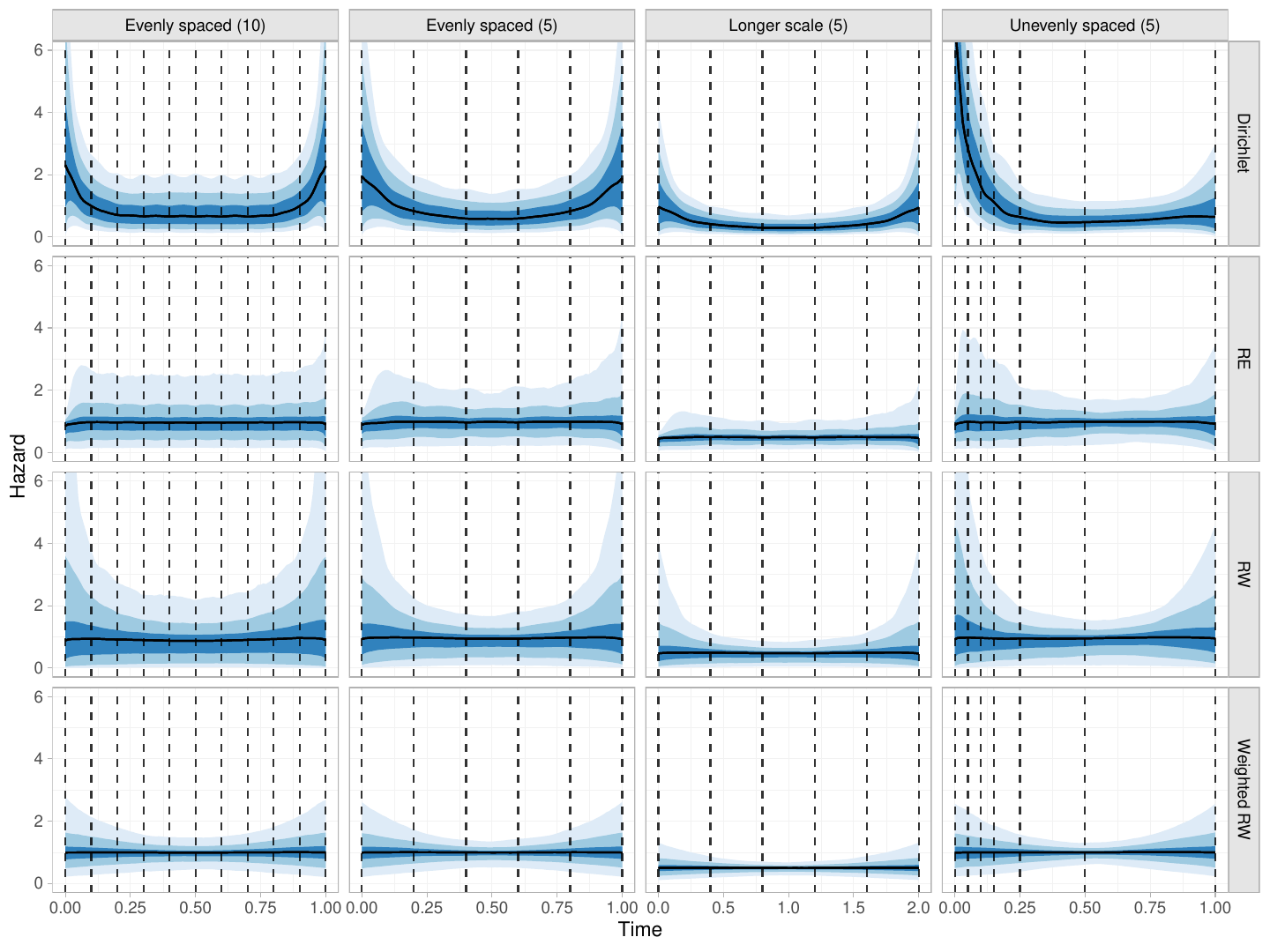}
  \caption{Prior distributions of the baseline hazards implied by $\operatorname{Dirichlet}(\bm{1})$, random effect (RE), standard and weighted random walk (RW) priors on the (inverse-softmax) spline coefficients of a cubic M-spline, for different knot placements and timescales. The solid line is the prior median; shaded ribbons indicate 50\%, 80\%, and 95\% credible regions of the prior density. Vertical dashed lines indicate the knot locations. A weakly-informative prior on the standard deviation $\sigma \sim \Normal(0, 1^2)$ was used for the random effect and both random walk priors.}
  \label{fig:prior_basehaz_cubic}
\end{figure}

\Cref{fig:prior_basehaz_cubic} illustrates the prior distributions on the baseline hazard implied by different types of prior distribution for the spline coefficients.
The unevenly spaced knots in the rightmost column are representative of practical applications, where knot locations are derived from quantiles of the observed event times: this leads to most knots being placed early in follow-up where there are more individuals and so more events.
The $\operatorname{Dirichlet}(\bm{1})$ prior distribution on $\bm{\alpha}_j$ is uniform over all possible coefficient vectors (i.e.\ the unit simplex) and does not lead to any shrinkage.
Moreover, we see that the prior baseline hazard implied by $\operatorname{Dirichlet}(\bm{1})$ is not constant and is heavily dependent on the location of the knots, increasing markedly in regions with more knots.
The random effect prior on $\bm{\alpha}^*_j$ centered around a constant baseline hazard, proposed by \citet{Jackson2023}, does lead to some shrinkage but not a sufficient amount to avoid overfitting as the number of knots increases.
Examining \cref{fig:prior_basehaz_cubic}, we see that this is at least in part due to the dependence of the implied prior baseline hazard on the number and location of knots: the prior variance increases in regions with more knots, which counteracts the effect of shrinkage and leads to under-smoothing.
An unweighted random walk prior on $\bm{\alpha}^*_j$, obtained by setting all $w_{j,l}=1$ in \cref{eqn:rw1_prior}, has the same problem; again the prior variance is increased in regions with more knots, leading to under-smoothing.
We see that the weighted random walk prior on $\bm{\alpha}^*_j$ that we propose resolves these issues and results in a prior baseline hazard that is invariant to the number or location of the knots.

\subsection{Modelling non-proportional hazards}\label{sec:methods_nph}

Whilst the M-spline model proposed in \eqref{eqn:mspline_ph} allows the shape of baseline hazards to be modelled very flexibly, this is still a proportional hazards model: the same baseline hazard is used for every arm within each trial.
Here, we propose two ways that this assumption may be relaxed: stratifying the baseline hazard by treatment arm, and introducing treatment effects onto the spline coefficients.

\subsubsection{Stratified baseline hazards}

The simplest way to relax the proportional hazards assumption is to further stratify the baseline hazards by treatment arm, as well as by study, which can be written as
\begin{subequations}\label{eqn:mspline_nph_stratified}
\begin{align}
	S_{jk}(t) &= \exp\mleft( -\bm{\alpha}_{jk}\tr \bm{I}_\kappa(t, \bm{\zeta}_j) \exp(\eta_{jk}) \mright)\\
	h_{jk}(t) &= \bm{\alpha}_{jk}\tr \bm{M}_\kappa(t, \bm{\zeta}_j) \exp(\eta_{jk}).
\end{align}
\end{subequations}
The only difference between this model and the proportional hazards model \eqref{eqn:mspline_ph} is that the spline coefficients $\bm{\alpha}_{jk}$ are now defined to be independent for each treatment arm in each study, as opposed to common for all arms of a study $\bm{\alpha}_j$, and are each given an independent random walk prior distribution.

This stratified model \eqref{eqn:mspline_nph_stratified} relaxes the proportional hazards assumption in a general manner, without requiring any additional modelling assumptions.
Whilst this model may be useful for assessing the proportional hazards assumption, it is of little use as a final model for decision-making, as predictions can only be made for observed treatment arms in each study population in the network.

\subsubsection{Treatment effects on spline coefficients}

Producing predictions on any treatment in any target population requires departures from non-proportionality to be modelled.
We propose the following M-spline model where treatment effects are introduced onto the inverse-softmax transformed spline coefficients
\begin{subequations}\label{eqn:mspline_nph_regression}
\begin{align}
	S_{jk}(t) &= \exp\mleft( -\bm{\alpha}_{jk}\tr \bm{I}_\kappa(t, \bm{\zeta}) \exp(\eta_{jk}) \mright) \\
	h_{jk}(t) &= \bm{\alpha}_{jk}\tr \bm{M}_\kappa(t, \bm{\zeta}) \exp(\eta_{jk}) \\
	\operatorname{softmax}^{-1}(\bm{\alpha}_{jk}) &= \bm{\alpha}^*_{j} + \bm{\gamma}^{(\alpha)}_{k}
\end{align}
\end{subequations}
where $\bm{\gamma}^{(\alpha)}_k$ is a vector of length $L + \kappa - 1$ for the non-proportional effect of the $k$-th treatment on each of the spline coefficients.
Under this model, the $\bm{\alpha}^*_{j}$ are interpreted as study-specific intercepts for the linear predictor on the inverse-softmax spline coefficients; the proportional hazards model \eqref{eqn:mspline_ph} is equivalent to an ``intercept only'' model on the spline coefficients.
As before, we place a weighted random walk prior distribution on $\bm{\alpha}^*_{j}$, centered around a constant baseline hazard.
However, for this model we must use the same vector of knot locations for every study in the network, $\bm{\zeta}_j = \bm{\zeta}$ for all $j$, and thus the random walk prior means $\bm{\varphi}_j = \bm{\varphi}$ and weights $\bm{w}_j = \bm{w}$ are also common across all studies $j$.

The common knot vector $\bm{\zeta}$ may in principle be chosen in any manner, however placing a knot just before the last observation time in a study is likely to lead to difficulties in estimation.
Some reasonable choices are to use quantiles of the observed event times in the study with the longest follow-up, to use quantiles of the observed event times in all studies lumped together, or to take quantiles of the observed event times in each study separately and then use the quantiles of these combined across studies.
We choose the latter as a reasonable default, as this appears in practice to be most likely to produce well-behaved knot locations when studies have varying lengths of follow-up.
Nevertheless, we recommend visualising the knot locations overlaid on Kaplan-Meier curves for each study and making adjustments as necessary, prior to model fitting.

We handle the $K$ non-proportionality effect vectors $\bm{\gamma}^{(\alpha)}_k$ symmetrically, with a weighted multivariate random walk prior distribution centered around zero (no departure from proportionality):
\begin{subequations}\label{eqn:nph_rw1_prior}
\begin{align}
	\mleft( \gamma^{(\alpha)}_{l,1}, \dots, \gamma^{(\alpha)}_{l,K} \mright)\tr	&= \sum_{m = 1}^l \bm{v}_m \quad \forall l=1,\dots,L+\kappa-1\\
	\bm{v}_l &\sim \MVN \mleft( 0, w_l \bm{\sigma}^{(\alpha)\tr} \bm{P} \bm{\sigma}^{(\alpha)} \mright) \quad \forall l=1,\dots,L+\kappa-1
\end{align}
\end{subequations}
where $\bm{P}$ is a $K \times K$ correlation matrix and $\bm{\sigma}^{(\alpha)} = \mleft( \sigma^{(\alpha)}_1, \dots, \sigma^{(\alpha)}_K \mright)\tr$ is a vector of smoothing standard deviations.
We choose to make the simplifying assumption of a common smoothing standard deviation $\sigma^{(\alpha)}_k = \sigma^{(\alpha)}$ for all treatments, which implies that $\bm{P}$ has 0.5 correlations in every off-diagonal element, and we choose $\sigma^{(\alpha)} \sim \halfNormal(0, 1^2)$ as a weakly-informative prior distribution.
As the smoothing standard deviation $\sigma^{(\alpha)}$ approaches zero, this model approaches a proportional hazards model.

This weighted multivariate random walk prior distribution \eqref{eqn:nph_rw1_prior} smooths departures from proportionality over time and induces shrinkage to avoid overfitting.
This approach handles each of the $K$ treatments symetrically; each treatment arm in the network has the same prior variation.
If instead we set $\bm{\gamma}^{(\alpha)}_1 = 0$, then all study arms for the reference treatment in the network are forced to have lower variation (both prior and posterior) than the other treatment arms, and we do not have a symmetric model; changing the reference treatment results in substantially different estimates.
We can translate the symmetric non-proportionality effects $\bm{\gamma}^{(\alpha)}_{k}$ into contrasts for the effect of non-proportionality for each treatment versus the network reference treatment as $\bm{d}^{(\alpha)}_{k} = \bm{\gamma}^{(\alpha)}_k - \bm{\gamma}^{(\alpha)}_1$.
These $\bm{d}^{(\alpha)}_{k}$ can then be interpreted as the relative departure from proportionality for treatment $k$ compared to treatment 1.

Here we only consider the non-proportionality effects $\bm{\gamma}^{(\alpha)}_{k}$ on the spline coefficients as fixed effects, so the departure from proportionality for each treatment is assumed to be the same across all studies.
In theory, a random effects model could also be specified on these parameters, allowing study-specific departures from proportionality to follow a random effects distribution. 
However, such a model is likely to be hard to estimate, and any such heterogeneity would also be captured by the typical random effects model on the main treatment effects $d_k$.

Inconsistency in the non-proportionality effects may be assessed by comparing this model to the stratified non-proportional hazards model \eqref{eqn:mspline_nph_stratified}, which is analogous to an unrelated mean effects model as it imposes no consistency relationships on the non-proportionality effects.
Similarly to the unrelated mean effects model, inconsistency is indicated by improvements in model fit for the stratified non-proportional hazards model over the non-proportional hazards model with treatment effects on the spline coefficients, either overall or locally for a study or studies, and/or by a reduction in the between-study heterogeneity $\tau$ for random effects models.
However, since the baseline hazards in the stratified non-proportional hazards model are completely independent between studies this approach may also detect heterogeneity in the non-proportionality effects, and cannot necessarily distinguish between heterogeneity and inconsistency.

\subsection{Including covariates}

When baseline covariate information is available, the above models can be extended to incorporate covariate effects on the main log hazard rate linear predictor and/or on the spline coefficients to model departures from proportionality.
Let $\bm{x}_{ijk}$ be a vector of covariates for each individual $i$ in study $j$ receiving treatment $k$; study- or arm-level covariates are a special case where every constituent individual event time has the same covariate value.

With covariates, the log hazard rate linear predictor $\eta_{ijk}$ becomes
\begin{equation}
	\eta_{jk}(\bm{x}_{ijk}) = \mu_j + \bm{x}_{ijk}\tr \mleft( \bm{\beta}_1 + \bm{\beta}_{2,k} \mright) + \delta_{jk}
\end{equation}
where $\bm{\beta}_1$ and $\bm{\beta}_{2,k}$ are regression coefficients for prognostic and effect modifying covariates, respectively, and we set $\bm{\beta}_{2,1} = 0$ for the reference treatment.

The stratified non-proportional hazards model \eqref{eqn:mspline_nph_stratified} may be modified to stratify the spline coefficients by discrete covariates, as well as or instead of stratifying by treatment, denoted $\bm{\alpha}_{jk;\bm{x}_{ijk}}$ or $\bm{\alpha}_{j;\bm{x}_{ijk}}$.
If the spline coefficients are not stratified by treatment, then the resulting model can be used to make predictions for all treatments in every study population in the network, unlike models that stratify by treatment arm.

The non-proportional hazards model with treatment effects on the spline coefficients \eqref{eqn:mspline_nph_regression} may be extended to include covariate effects on the spline coefficients:
\begin{equation}
	\operatorname{softmax}^{-1}\mleft( \bm{\alpha}_{jk}(\bm{x}_{ijk}) \mright) = \bm{\alpha}^*_{j} + \mleft( \bm{B}^{(\alpha)}_1 + \bm{B}^{(\alpha)}_{2,k} \mright)\tr \bm{x}_{ijk} + \bm{\gamma}^{(\alpha)}_{k}
\end{equation}
where $\bm{B}^{(\alpha)}_1$ and $\bm{B}^{(\alpha)}_{2,k}$ are matrices of main covariate effects and treatment-covariate interactions on each of the (inverse-softmax) spline coefficients.
Each covariate is given an independent zero-mean random walk prior distribution, to smooth the effects on the departure from proportionality over time.
The non-proportional effects of each treatment $\bm{\gamma}^{(\alpha)}_{k}$ may also be omitted from this model, in which case departures from non-proportionality are modelled only in terms of the covariates.

\section{Case study: application to non-small cell lung cancer}

We now apply these methods to the non-small cell lung cancer example.
Analyses were carried out in R version 4.4.1 \citep{RCore} and Stan version 2.33.1 \citep{Carpenter2017}, using the \textit{multinma} R package \citep{multinma} version 0.7.2.
Analysis code and data are available from the GitHub repository \href{https://github.com/dmphillippo/mspline-survival-paper}{dmphillippo/mspline-survival-paper}.

First, we consider a cubic M-spline model that incorporates non-proportional hazards via treatment effects on the spline coefficients \eqref{eqn:mspline_nph_regression}.
Since this model requires a common vector of knot locations across all studies, we begin by visualising the default knot locations for seven internal knots chosen according to quantiles of the quantiles of observed event times in each study, as shown in \cref{fig:nsclc_knots_default} overlaid on the Kaplan-Meier curves from each study.
The final internal knot at around 20 weeks lies just before the end of follow-up in both the IMpower150 and PRONOUNCE studies, which is likely to lead to sampling issues.
We therefore choose to add an additional knot at the end of follow-up in the ERACLE study, shown in \cref{fig:nsclc_knots_modified}.

\begin{figure}
	\centering
  \includegraphics[width=\textwidth]{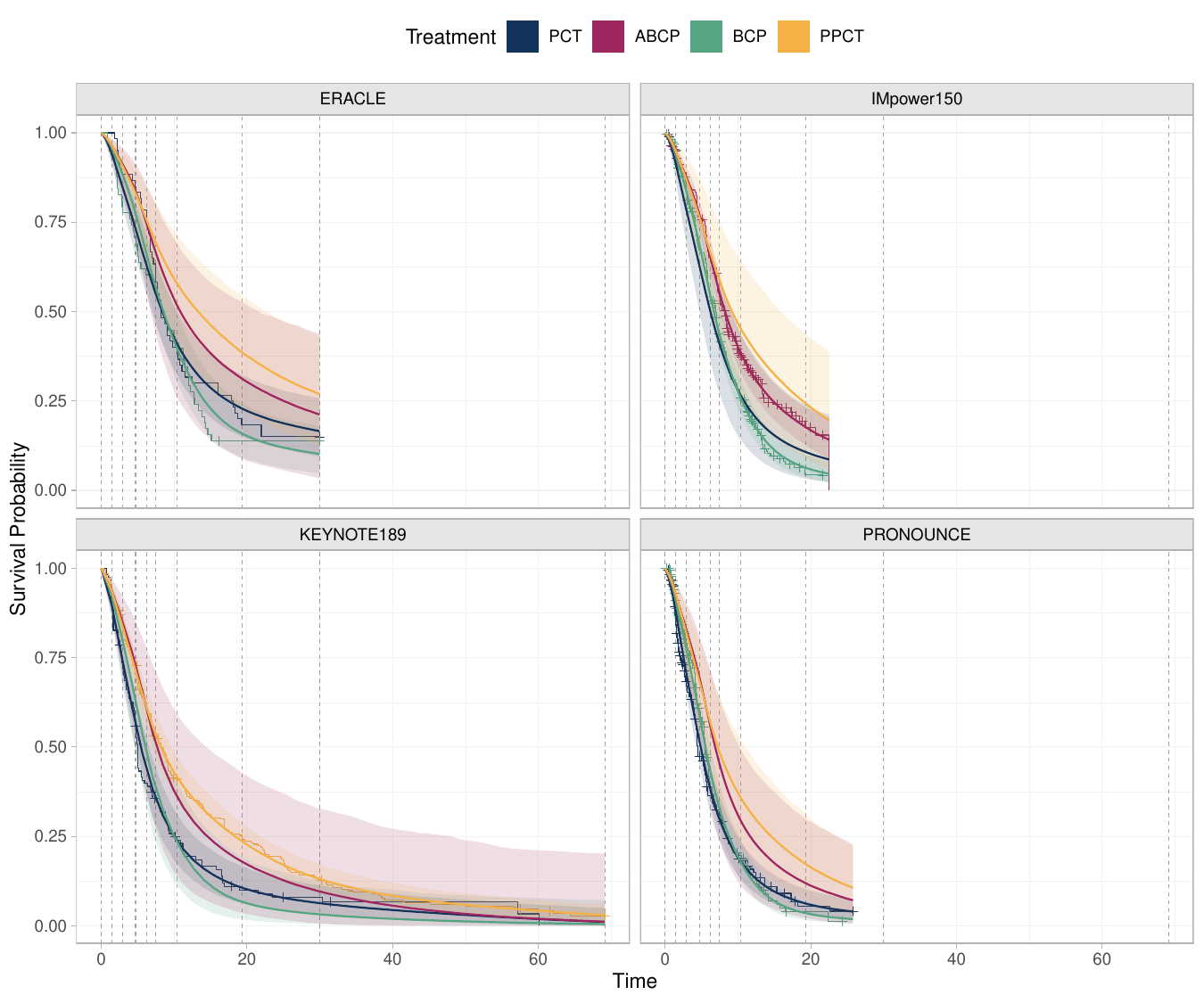}
  \caption{Estimated progression-free survival curves on each treatment, in each study population, from a cubic M-spline model with non-proportional hazards via treatment effects on the spline coefficients. Shaded bands indicate 95\% credible intervals. Observed Kaplan-Meier curves are also shown. Dashed vertical lines indicate the location of knots.}
  \label{fig:nsclc_survival_nph}
\end{figure}

After fitting this model, the estimated survival curves are shown in \cref{fig:nsclc_survival_nph} and the estimated baseline hazards are shown in \cref{fig:nsclc_hazard_nph}.
The estimated survival curves are a good visual fit to the data. 
For comparison, we also fit a proportional hazards model \eqref{eqn:mspline_ph} and a non-proportional hazards model with stratified baseline hazards \eqref{eqn:mspline_nph_stratified}.
The proportional hazards model is clearly inappropriate here, as the estimated survival curves (\cref{fig:nsclc_survival_ph}) do not capture the crossing of survival curves in the ERACLE and PRONOUNCE studies, or the more subtle departures from non-proportionality in the KEYNOTE189 and IMpower150 studies.
The non-proportional hazards model with stratified baseline hazards provides good visual fit and captures the non-proportionality (\cref{fig:nsclc_survival_nph_stratified}), but can only produce estimates for the observed treatment arms in each study population.
Notably, the non-proportional hazards model with treatment effects on the spline coefficients produces very similar estimates to the non-proportional hazards model with stratified baseline hazards for the observed treatment arms, whilst also allowing predictions for the unobserved treatment arms.

We assess model fit using the leave-one-out information criterion (LOOIC) \parencite{Vehtari2016}.
Model fit statistics for each of the models are presented in \cref{tab:ns1_looic}.
The non-proportional hazards model with treatment effects on the spline coefficients provides the best overall fit with the lowest LOOIC (7470.8), followed by the non-proportional hazards model with stratified baseline hazards (7489.8), and then the proportional hazards model is the worst fit (7493.1).
However, it is important to also assess the model fit within each study in the network, as there may be differences in model fit in individual studies that are masked when looking at overall fit.
The non-proportional hazards model with treatment effects on the spline coefficients has the lowest LOOIC within each study, with the exception of ERACLE which is best fit by the non-proportional hazards model with stratified baseline hazards.
This indicates the presence of inconsistency or heterogeneity in the non-proportionality effects for the ERACLE study compared to the other studies in the network; here this is heterogeneity, as there are no loops of evidence.
Examining the Kaplan-Meier curves and estimated survival curves, we see that initial PFS until around week 7 on BCP and PCT treatments in ERACLE displays different behaviour to that in PRONOUNCE; in ERACLE the PCT arm initially has better PFS than BCP, whereas in PROSPECT this pattern is reversed.
The non-proportional hazards model with stratified baseline hazards is free to fit both patterns of initial PFS separately (\cref{fig:nsclc_survival_nph_stratified}), whereas the non-proportional hazards model with treatment effects on the spline coefficients synthesises the observed relationships and results in a fit that is closer to the larger PROSPECT study (\cref{fig:nsclc_survival_nph}).

To assess the adequacy of the number of knots, we also fit a model with a larger number of knots (11 internal knots as opposed to 8).
The LOOIC model fit statistics are not substantially different, both overall and within each individual study (\cref{tab:ns1_looic}), and the estimated survival curves are unchanged (\cref{fig:nsclc_survival_nph_11kt}).
This suggests that the initial choice of the number of knots (8) is sufficient, and demonstrates the shrinkage behaviour of the random walk prior distributions that prevent overfitting as the number of knots increases.

\begin{figure}
	\centering
  \includegraphics[width=\textwidth]{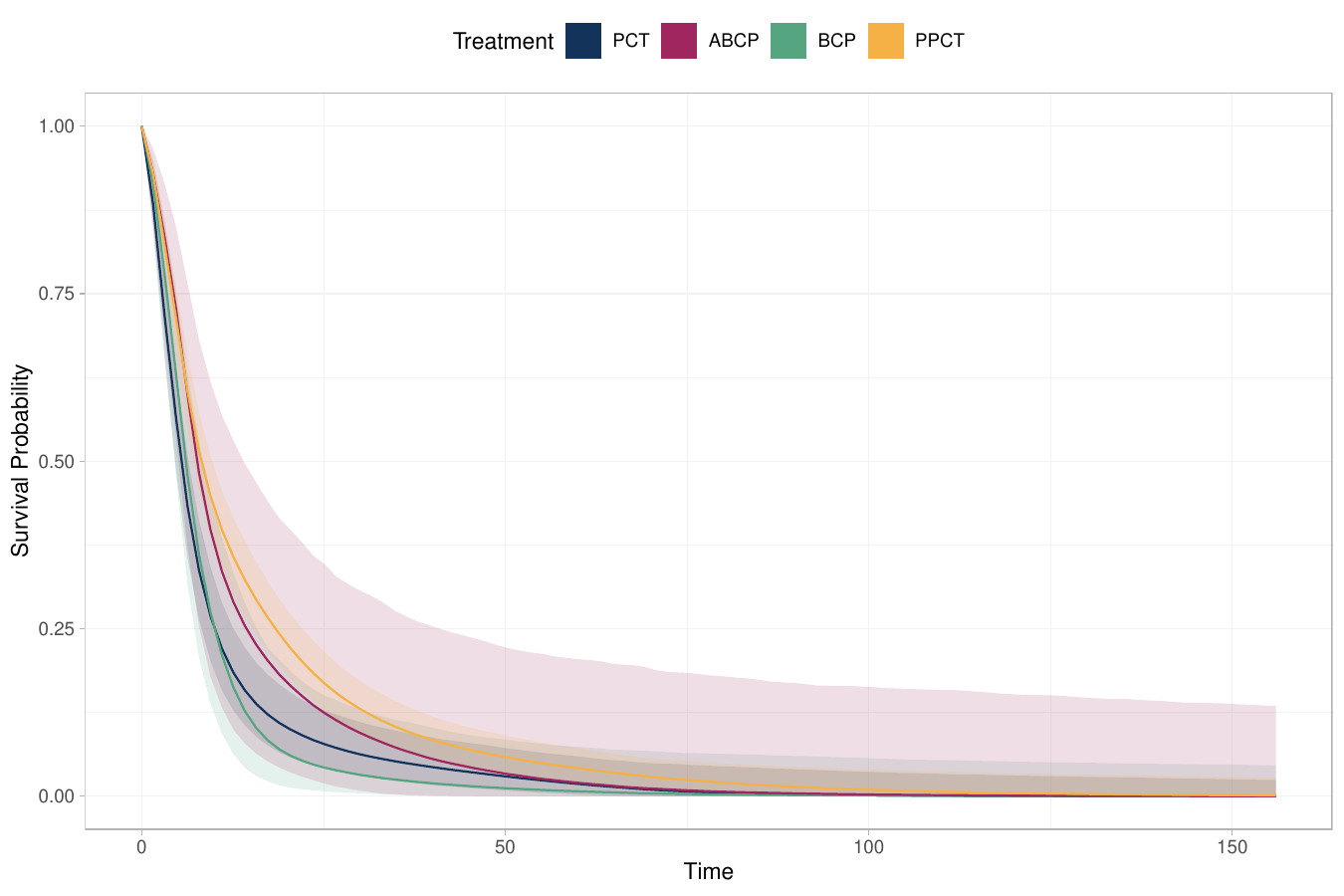}
  \caption{Estimated progression-free survival curves on each treatment in the KEYNOTE189 population, from a cubic M-spline model with non-proportional hazards via treatment effects on the spline coefficients. Shaded bands indicate 95\% credible intervals.}
  \label{fig:nsclc_keynote_survival}
\end{figure}

Finally, we produce estimates for all treatments using the baseline hazard from the KEYNOTE189 population, which was deemed representative for decision-making purposes.
\Cref{fig:nsclc_keynote_survival} shows the estimated survival curves on each treatment in the KEYNOTE189 population, extrapolated out to three years with a constant hazard after the final knot at the end of follow-up (69 months).
The time-varying marginal log hazard ratios for each treatment compared to PCT are shown in \cref{fig:nsclc_keynote_hr}.
Note the ABCP and BCP treatments are only observed up until 22.5 and 30 months, respectively, after which the model shrinks back towards a proportional hazards model for these treatments.

\section{Discussion}\label{sec:gl_discussion}


In this paper, we proposed a novel approach to network meta-analysis of time-to-event outcomes using M-splines to flexibly model the baseline hazard.
Other flexible models include the Royston-Parmar model \parencite{Freeman2017,Royston2002} and fractional polynomials \parencite{Jansen2011}; however, the M-spline approach presented here has several practical advantages over these alternatives.
Firstly, whilst described in a Bayesian setting for NMA by \textcite{Freeman2017}, the Royston-Parmar model is intractable for Bayesian analysis as the cumulative hazard is not constrained to be monotonically increasing \parencite{Royston2002}, and is thus rarely used in practice.
The M-spline model by design ensures that the cumulative hazard is monotonically increasing, and can be written in closed form in terms of I-splines.
Fractional polynomial models are widely used \parencite{Jansen2011}, however the process of fitting and comparing multiple models to select appropriate powers for modelling the baseline hazard is time-consuming.
By contrast, the M-spline model with the proposed random walk prior distribution allows the analyst to simply fit a single model with a ``large enough'' number of knots, allowing the model to shrink to an appropriate level of complexity.
Fractional polynomial models are also constrained to use the same powers across all treatments and studies in the network, which may be overly-restrictive if baseline hazard shapes differ substantially, whereas the M-spline model allows the baseline hazard to take entirely different shapes where necessary.

The M-spline and fractional polynomial models also differ in terms of how non-proportional hazards are modelled and how extrapolation is handled.
Both models have ``main'' treatment effects, conditional log hazard ratios acting on the hazard rate, which in the M-spline model are $d_k$.
However, the treatment effects (and corresponding additivity assumptions and consistency equations) for departures from non-proportionality are applied on different scales: in the M-spline model \eqref{eqn:mspline_nph_regression} these treatment effects $\bm{d}^{(\alpha)}_k$ are placed on the inverse-softmax transformed spline coefficients; in the fractional polynomial model these are placed on the fractional polynomial coefficients on the log hazard scale.
Moreover, the two models represent fundamentally different approaches to modelling departures from proportional hazards.
The M-spline model estimates local deviations from proportional hazards for each treatment within each time period defined by the knots, smoothed over time, whereas the fractional polynomial model estimates global deviations over the whole timescale as mixtures of polynomial terms.
The two approaches also behave differently when extrapolating past the end of follow-up: the fractional polynomial model continues the polynomial fit \emph{ad infinitum}, which can result in unrealistic extrapolations, whereas the M-spline model shrinks towards constant proportional hazards.
When external long-term data are available, for example from a disease registry or cohort study, these may be used to inform long-term survival estimates within the M-spline model \parencite{Jackson2023}.
Since the fit between two knots is only informed by data up to $\kappa-1$ knots later (\cref{sec:mspline_basis}), with this approach the early part of the survival curve is entirely estimated from RCT evidence and is not affected by long-term data and vice versa, with smooth interpolation in-between.
The M-spline model may also be extended to incorporate different extrapolation assumptions on the long-term effect of treatment such as waning or cure \parencite{Jackson2023}.

In the non-small cell lung cancer case study, we fitted a flexible non-proportional hazards model with treatment effects on the spline coefficients to model departures from proportionality.
In this example, we only had access to outcome times and censoring indicators in each study, reconstructed from digitised Kaplan-Meier curves.
However, we have also described how baseline covariates can be incorporated into regressions on the main log hazard rate linear predictor and/or on the spline coefficients.
If individual-level covariates are available from every study then the resulting analysis is a one-stage individual participant data network meta-regression.
When individual participant data are only available for a subset of studies in the network, with published summary statistics on baseline covariates and (reconstructed) outcome times and censoring indicators available from the rest, these may be combined in a coherent manner using multilevel network meta-regression, as described and demonstrated by \textcite{Phillippo2024_survival}.
Adjusting for covariates can account for differences between study populations, reducing heterogeneity and inconsistency, and removing bias due to imbalances in effect-modifying covariates between populations.
Estimates can then be produced for a given target population of interest; this is particularly crucial in decision-making contexts such as health technology assessment, where such analyses are increasingly common and referred to as ``population adjustment'' \parencite{TSD18,Phillippo2024_survival}.
Adjusting for individual-level covariates can also reduce or even remove issues of non-proportional hazards, since omitted covariates are one potential cause of non-proportionality at the marginal level \parencite{Therneau2000,Phillippo2024_survival}.

Due to the small size of the non-small cell lung cancer network we fitted only fixed effect models, and there was no possibility of inconsistency as there were no loops of evidence.
However, in general, heterogeneity and inconsistency must be assessed wherever possible, as with all forms of network meta-analysis.
We have described random effects models for incorporating heterogeneity, and unrelated mean effects and node-splitting models for assessing inconsistency, all of which are implemented in the \emph{multinma} R package \parencite{multinma}.
When non-proportional hazards are modelled through treatment effects on the spline coefficients \eqref{eqn:mspline_nph_regression}, there is also the potential for heterogeneity and inconsistency in these effects.
Heterogeneity in the non-proportionality effects could in theory be assessed by placing random effects on $\bm{\gamma}_k^{(\alpha)}$ to allow study-specific departures from non-proportionality.
This would require the use of a hierarchical correlated multivariate random walk, extending the multivariate random walk prior distribution \eqref{eqn:nph_rw1_prior} to incorporate between-study variation and within-study correlation.
However, such a model may be difficult to estimate in practice.
Moreover, heterogeneity in the non-proportionality effects will also be captured in random effects on the main treatment effects $d_k$.
We have suggested assessing inconsistency in the non-proportionality effects by comparing model fit (and between-study heterogeneity $\tau$ for random effects models) against the non-proportional hazards model with stratified baseline hazards \eqref{eqn:mspline_nph_stratified}.
However, this approach may also detect heterogeneity in the non-proportionality effects and cannot necessarily distinguish between these two types of violations of the same exchangeability assumption; in practice the distinction is unimportant, as long as violations are detected.
More standard inconsistency models for the non-proportionality effects are not straightforward to define due to the use of the symmetric model.
Both node-splitting models and unrelated mean effects models could be implemented under a non-symmetric model, but this results in substantially different estimates depending on the choice of the reference treatment, which is undesirable.
In contrast, the fractional polynomial approach can readily incorporate random effects and directly assess inconsistency on any of the treatment effect parameters, not just the main treatment effects on the hazard rate \parencite{Jansen2011}, although this is rarely done in practice.

Fitting M-splines to the baseline hazard requires the specification of a vector of knot locations.
Due to the use of random walk priors that encourage shrinkage, the results are not typically sensitive to the choice of knot locations provided that a sufficiently large number of knots are chosen.
For the proportional hazards model and the non-proportional hazards model with stratified baseline hazards, each study may have a separate vector of knot locations and the default choice (evenly-spaced quantiles of the observed event times) works well for all models.
However, for the non-proportional hazards model with treatment effects on the spline coefficients a common knot vector must be used for all studies.
This complicates knot placement, especially when studies have very different lengths of follow-up, since placing a knot just before the end of follow-up in a study is likely to lead to estimation difficulties.
The simple quantile-based knot locations that we have proposed work well when studies all have similar lengths of follow-up, but can sometimes provide knot locations that frustrate estimation when studies have very different lengths of follow-up. 
We therefore suggest to check the generated knot locations prior to model fitting, and to modify knot placement if there are any intervals where the final observations occur just after a knot.
Future research could develop knot placement algorithms that better handle studies having varied lengths of follow-up and avoid placing knots at locations likely to lead to sampling difficulties.

In decision-making contexts such as health technology assessment, estimated survival curves on each treatment in a representative target population are used as inputs to an economic model, in order to evaluate cost-effectiveness.
Despite the limitations, the majority of such economic models are still built in Excel \parencite{Incerti2019}.
Samples from the estimated survival curves over time cannot typically be used directly due to computational issues, so instead standard parametric and fractional polynomial models are often evaluated within Excel by sampling from (multivariate) Normal approximations for the parameters.
A similar approach may be used to evaluate M-spline survival curves in Excel.
A multivariate Normal approximation may be made for the inverse-softmax spline coefficients $\operatorname{softmax}^{-1}(\bm{\alpha}_{jk})$, i.e.\ for $\bm{\alpha}^*_j$ for a proportional hazards model, or for $\bm{\alpha}^*_j + \bm{\gamma}^{(\alpha)}_k$ for the non-proportional hazards model with treatment effects on the spline coefficients.
A multivariate Normal approximation is also made for the log hazard rate linear predictor on each treatment $\eta_{jk}$.
The integrated M-spline basis $\bm{I}_\kappa(t, \bm{\zeta}_j)$ is evaluated at a grid of time-steps $t$ and provided as data to Excel (e.g. as a matrix with number of columns equal to the dimension of the M-spline basis $L+\kappa$, and number of rows equal to the number of time-steps $t$).
The survival curves $S_{jk}(t)$ on each treatment can then be evaluated within Excel using these inputs following equation \eqref{eqn:mspline_ph_S}.
For economic models built in R, Python, Julia or other suitable statistical software, posterior samples from the survival curves at each time-step can be used directly and Normal approximations are not required.

The M-spline NMA model proposed in this paper is the first time that M-splines have been incorporated within the NMA framework.
This approach models baseline hazards in a flexible manner without restriction on the shape of the baseline hazard, and allows non-proportional hazards to be modelled through the effects of treatment and/or covariates on the spline coefficients.
Another key contribution is the novel random walk prior distribution that provides shrinkage and is invariant to the choice of knots or timescale, which is widely applicable to M-spline implementations in other settings beyond NMA.
These methods are implemented in the user-friendly \emph{multinma} R package \parencite{multinma}, which supports analysis of aggregate data, individual participant data, or mixtures of both in a multilevel network meta-regression \parencite{Phillippo2024_survival}, making the approach readily available to analysts.

\section*{Supplementary material}
Supplementary material is available online.

\section*{Funding}
DMP was supported by the UK Medical Research Council, grant number MR/W016648/1.
AS, HP, and NJW were supported by the UK National Institute for Health and Care Research (NIHR131974).
The views expressed are those of the authors and not necessarily those of the NIHR or the Department of Health and Social Care.

\section*{Conflicts of Interest}
HP is an employee of the University of Bristol, and employed as a statistical consultant for the pharmaceutical industry working for ConnectHEOR.
The remaining authors declare no conflicts of interest.

\section*{Data Availability}
The \emph{multinma} R package is freely available from CRAN at \url{https://cran.r-project.org/package=multinma}; see the package website \url{https://dmphillippo.github.io/multinma/} for further details. 
Code and data for the analyses presented in this paper are available from the GitHub repository \href{https://github.com/dmphillippo/mspline-survival-paper}{dmphillippo/mspline-survival-paper}.

\printbibliography[heading=bibintoc,title={Bibliography}]

\clearpage
\appendix
\setcounter{secnumdepth}{3}
\setcounter{equation}{0}
\setcounter{table}{0}
\setcounter{figure}{0}
\renewcommand{\theequation}{\Alph{section}.\arabic{equation}}
\renewcommand{\thetable}{\Alph{section}.\arabic{table}}
\renewcommand{\thefigure}{\Alph{section}.\arabic{figure}}
\crefalias{section}{appendix}
\crefalias{subsection}{appendix}

\pagenumbering{arabic}
\section{Appendix}
\subsection{Definition of M-spline basis}\label{sec:mspline_basis}

The M-spline and I-spline bases are constructed using the recursive formulae of \citet{Ramsay1988}, which are implemented in the \textit{splines2} R package \citep{splines2}.

Given a knot vector $\bm{\zeta}_j = (\zeta_{j,0}, \cdots, \zeta_{j, L+1})$, define an augmented knot vector $\bm{\zeta}_j^*$ of length $L+2\kappa$ by 
\begin{equation}\label{eqn:augmented_knots}
	\zeta^*_{j,s} = 
	\begin{cases}
		\zeta_{j,0} & \textrm{for } s = 1, \dots, \kappa \\
		\zeta_{j,s-\kappa} & \textrm{for } s = \kappa+1, \dots, \kappa+L \\
		\zeta_{j,L+1} & \textrm{for } s = \kappa+L+1, \dots, L+2\kappa
	\end{cases}
\end{equation}
In other words, the knot vector $\bm{\zeta}_j$ is padded by $\kappa$ replications of the lower and upper boundary knots at the start and end respectively to obtain the augmented knot vector $\bm{\zeta}^*_j$.
The M-spline basis $M_{\kappa, s}(t, \bm{\zeta}^*_j)$ at a given time $t$ for each dimension $s = 1,\dots,L+\kappa$ is then defined recursively by
\begin{subequations}
\begin{align}
	M_{1, s}(t, \bm{\zeta}^*_j) &= \frac{1}{\zeta^*_{j, s+1} - \zeta^*_{j, s}} \quad \textrm{if } \zeta^*_{j, s} \le t < \zeta^*_{j, s+1} \textrm{, otherwise } 0 \\
  M_{r, s}(t, \bm{\zeta}^*_j) &= \frac{r \mleft( (t - \zeta^*_{j,s}) M_{r-1, s}(t, \bm{\zeta}^*_j) + (\zeta^*_{j,s+1} - t) M_{r-1, s+1}(t, \bm{\zeta}^*_j) \mright)}{(r-1)(\zeta^*_{j, s+r} - \zeta^*_{j, s})}
\end{align}
\end{subequations}
for $r = 2,\dots,\kappa$.

We note two particular properties of this M-spline basis.
Firstly, $M_{\kappa,s}(t, \bm{\zeta}^*_j) > 0$ only for $\zeta^*_{j, s} \le t < \zeta^*_{j, s+\kappa}$ and is zero everywhere else. 
In other words, the spline fit between knots $\zeta_{j,s}$ and $\zeta_{j,s+1}$ is only informed by data up to $\kappa-1$ knots later, up to time $\zeta_{j,s+\kappa}$. 
Secondly, $\int_{\zeta_{j,0}}^{\zeta_{j,L+1}} \sum_{s=1}^{L+\kappa} M_{\kappa,s}(t,\bm{\zeta}^*_j) \dd t =1$, that is the M-spline is normalised to have integral 1 between the boundary knots.

The corresponding I-spline basis evaluated at time $t$ is the integral of the M-spline basis from the lower boundary point up until time $t$
\begin{equation}
	I_{\kappa,s}(t, \bm{\zeta}_j) = \int_{\zeta_{j,0}}^t M_{\kappa,s}(v, \bm{\zeta}_j) \dd v,
\end{equation}
which is a piecewise polynomial of degree $\kappa$.

\subsection{Modification of prior weights for a piecewise exponential hazard}\label{sec:rw_prior_pexp}

Degree-zero M-splines ($\kappa = 1$) correspond to a piecewise exponential baseline hazard.
However, the weights $\bm{w}_j$ are degenerate when $\kappa=1$, as equation \eqref{eqn:mspline_rw1_weights} results in zero divided by zero.
Instead, we propose to use the following modification of the weights for the piecewise exponential case:
\begin{equation}
	\bm{w}_j = \frac{\bm{\zeta}^*_{j,2:(L+1)} - \bm{\zeta}^*_{j,1:L}}{\zeta^*_{j,L+1} - \zeta^*_{j,1}}.
\end{equation}
In a similar fashion to before, these weights reflect the distance between each pair of knots and are normalised to sum to 1.

\begin{figure}
	\centering
  \includegraphics[width=\textwidth]{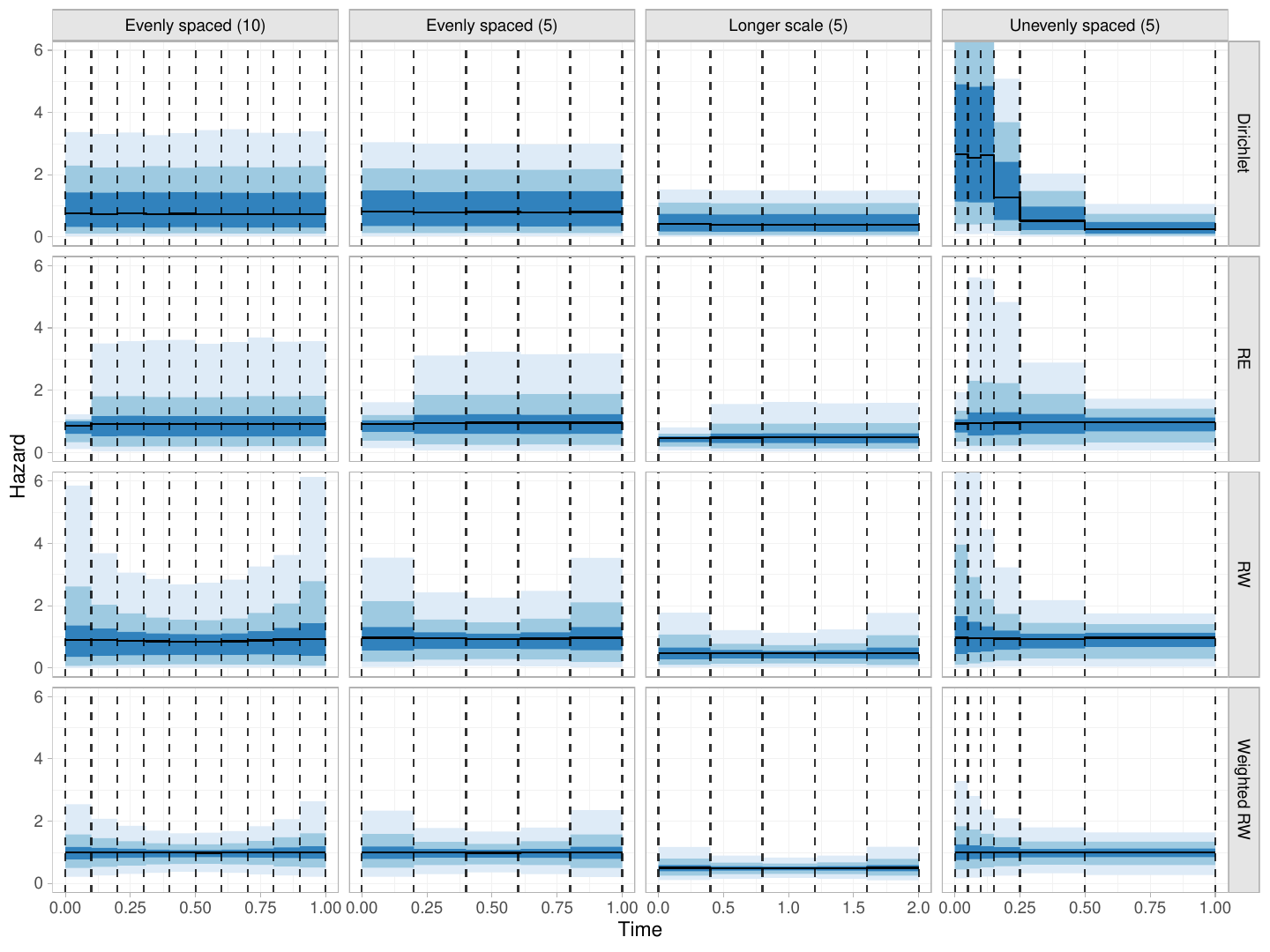}
  \caption{Prior distributions of the baseline hazards implied by Dirichlet, random effect (RE), standard and weighted random walk (RW) priors on the (inverse-softmax) spline coefficients of a degree-zero M-spline (i.e.\ piecewise exponential), for different knot placements and timescales. The solid line is the prior median; shaded ribbons indicate 50\%, 80\%, and 95\% credible regions of the prior density. Vertical dashed lines indicate the knot locations.}
  \label{fig:prior_basehaz_pexp}
\end{figure}

\Cref{fig:prior_basehaz_pexp} illustrates the prior distributions on the piecewise exponential baseline hazard implied by different types of prior distribution for the spline coefficients.
Here we see a similar story to that for the cubic M-spline baseline hazard in \cref{fig:prior_basehaz_cubic}.
The implied prior baseline hazard from a Dirichlet prior distribution on $\bm{\alpha}_j$ is not constant when the knots are unevenly spaced and increases markedly in regions with more knots.
The prior baseline hazard from both the random effect and unweighted random walk priors on $\bm{\alpha}^*_j$ depends on the number and location of knots; the prior variance increases in regions with more knots, which counteracts the effect of shrinkage and leads to under-smoothing.
The proposed weighted random walk prior on $\bm{\alpha}^*_j$ resolves these issues and results in a prior baseline hazard that is approximately invariant to the number or location of the knots; the invariance is approximate in this case, coarsened by the discrete nature of the piecewise exponential hazard.

\clearpage
\subsection{Additional tables and figures for NSCLC analysis}

\begin{figure}
	\centering
  \includegraphics[width=\textwidth]{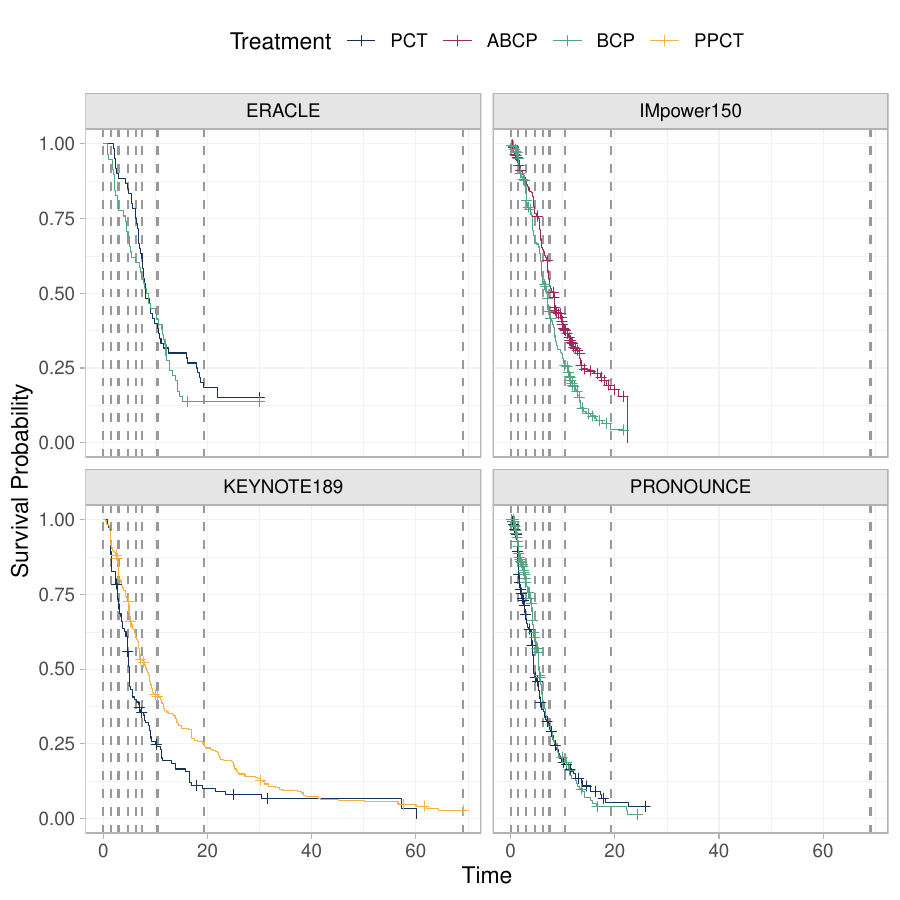}
  \caption{Default common knot locations overlaid on Kaplan-Meier curves in each study population. Seven internal knots are chosen according to quantiles of the quantiles of the observed event times in each study. Follow-up in the IMpower150 and PRONOUNCE studies ends just after the last internal knot, which is likely to lead to sampling issues.}
  \label{fig:nsclc_knots_default}
\end{figure}

\begin{figure}
	\centering
  \includegraphics[width=\textwidth]{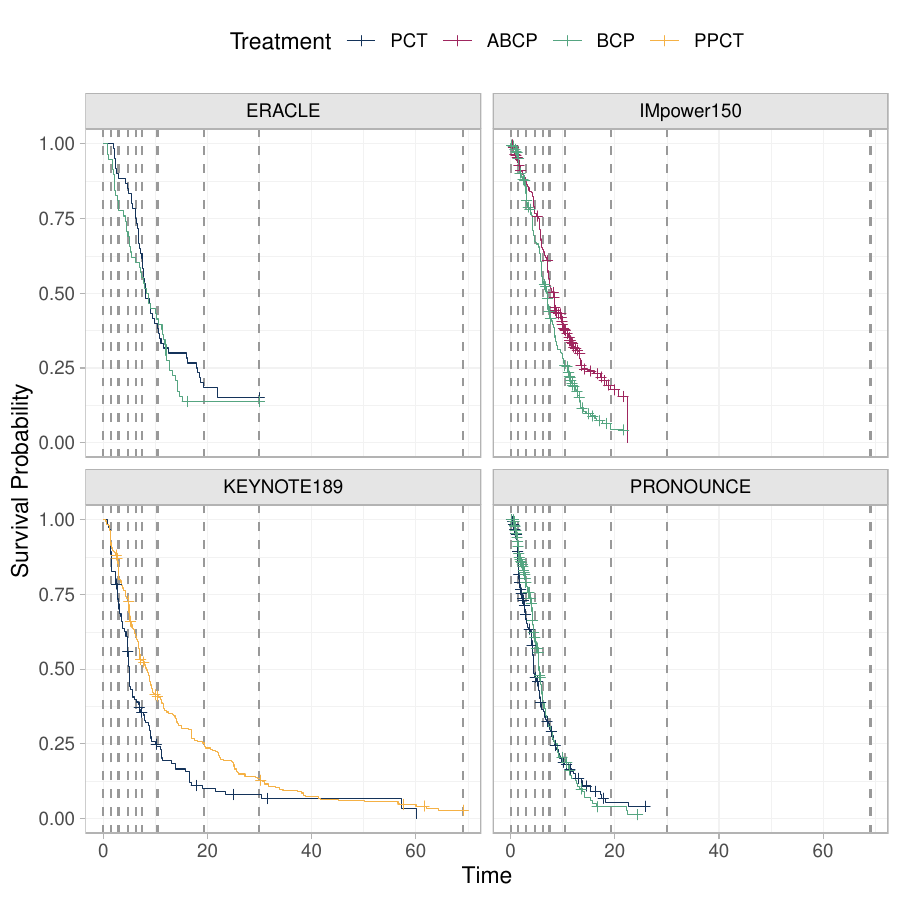}
  \caption{Modified knot locations, with an additional knot added at the end of follow-up of the ERACLE study.}
  \label{fig:nsclc_knots_modified}
\end{figure}

\begin{figure}
	\centering
  \includegraphics[width=\textwidth]{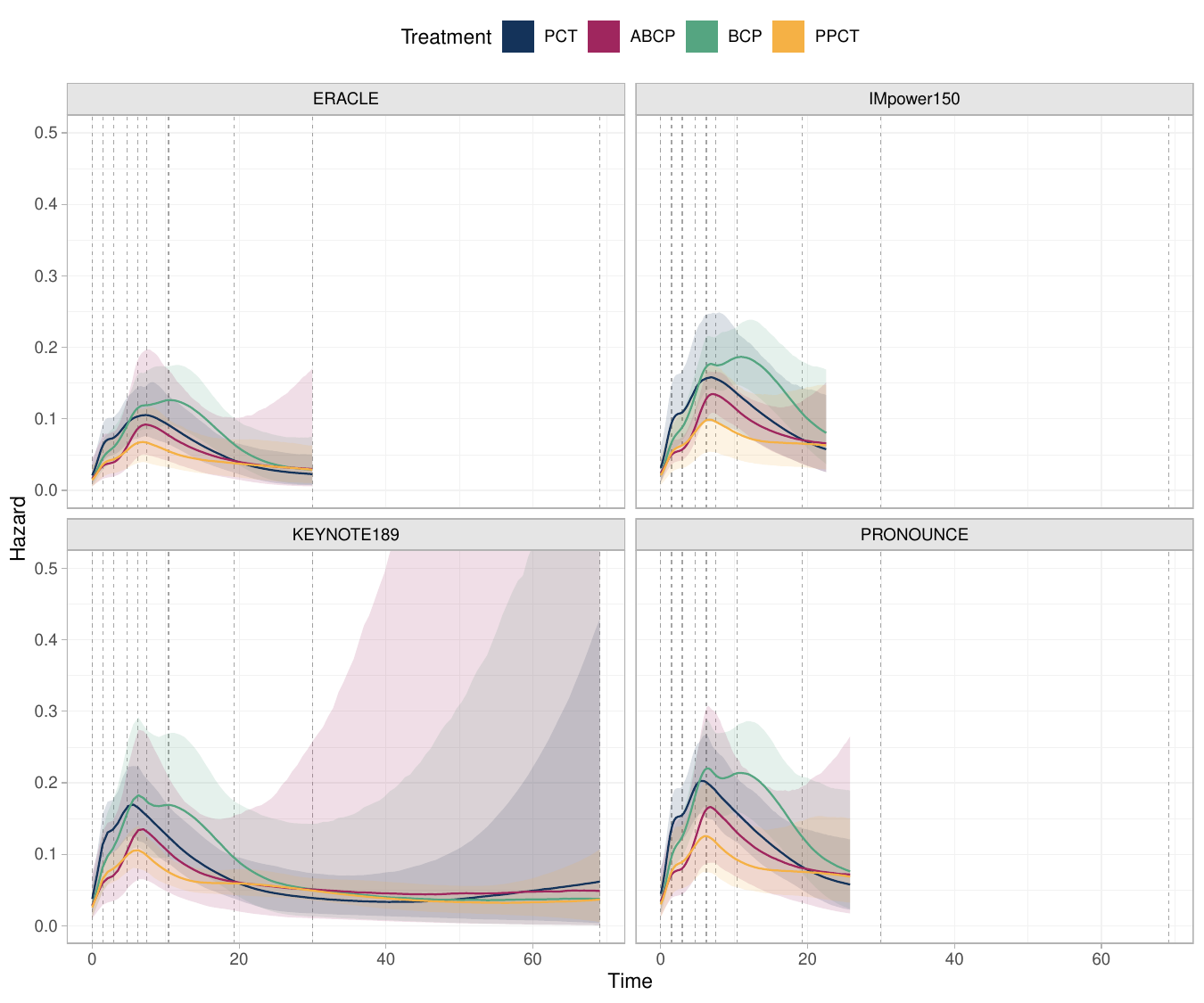}
  \caption{Estimated baseline hazards on each treatment in each study population, from a cubic M-spline model with non-proportional hazards via treatment effects on the spline coefficients. Shaded bands indicate 95\% credible intervals. Dashed vertical lines indicate the location of knots.}
  \label{fig:nsclc_hazard_nph}
\end{figure}

\begin{figure}
	\centering
  \includegraphics[width=\textwidth]{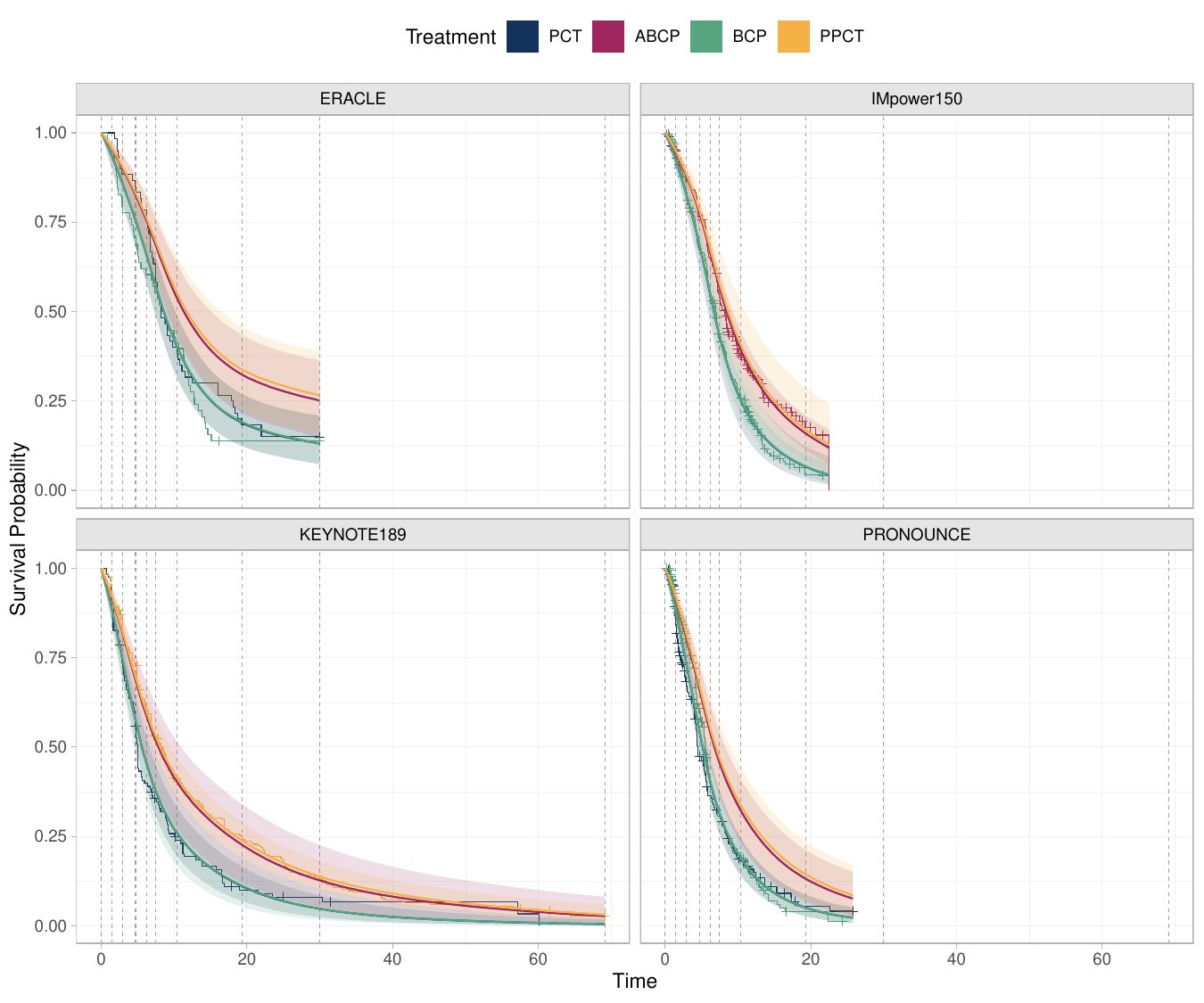}
  \caption{Estimated progression-free survival curves on each treatment in each study population, from a cubic M-spline model with proportional hazards. Shaded bands indicate 95\% credible intervals. Observed Kaplan-Meier curves are also shown. Dashed vertical lines indicate the location of knots.}
  \label{fig:nsclc_survival_ph}
\end{figure}

\begin{figure}
	\centering
  \includegraphics[width=\textwidth]{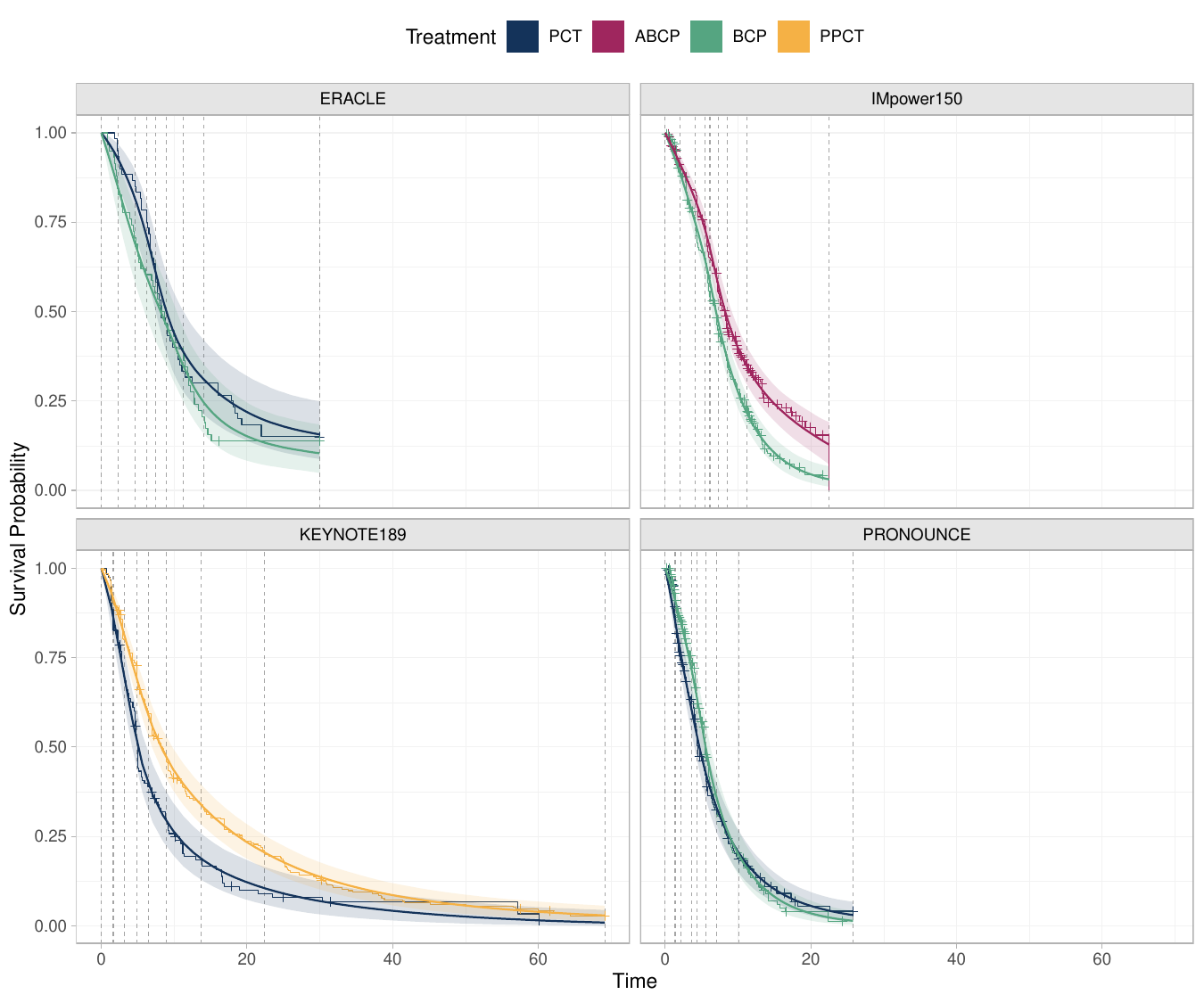}
  \caption{Estimated progression-free survival curves on each treatment in each study population, from a cubic M-spline model with non-proportional hazards via stratified baseline hazards for each treatment arm. Shaded bands indicate 95\% credible intervals. Observed Kaplan-Meier curves are also shown. Dashed vertical lines indicate the location of knots.}
  \label{fig:nsclc_survival_nph_stratified}
\end{figure}

\begin{table}[hp]
	\footnotesize
	\centering
	\caption{LOOIC model comparison statistics for ML-NMR models with and without the proportional hazards assumption, computed within each study in the network and overall. Lower values of LOOIC indicate better fit. $p_\mathrm{LOO}$ is an estimate of the effective number of model parameters.}

	\label{tab:ns1_looic}
	\begin{tabular}{lrrrr}
		\toprule
		& & \multicolumn{3}{c}{Non-proportional hazards} \\
		\cmidrule(l){3-5}
		& Proportional hazards & Treatment effects & Treatment effects & Stratified \\
		Study & 7 independent knots & 8 common knots & 11 common knots & 7 independent knots \\
		\midrule
		ERACLE     &  700.5 &  702.0 &  702.5 &  697.3 \\
		IMpower150 & 2744.8 & 2740.6 & 2741.8 & 2747.6 \\
		KEYNOTE189 & 2425.6 & 2418.9 & 2418.9 & 2425.1 \\
		PRONOUNCE  & 1622.3 & 1609.2 & 1609.3 & 1619.7 \\
		\midrule
		\textbf{Total} & \textbf{7493.1} & \textbf{7470.8} & \textbf{7472.5} & \textbf{7489.8} \\
		$p_\mathrm{LOO}$ & 21.9 & 27.5 & 30.7 & 32.0 \\
		\bottomrule
	\end{tabular}
\end{table}

\begin{figure}
	\centering
  \includegraphics[width=\textwidth]{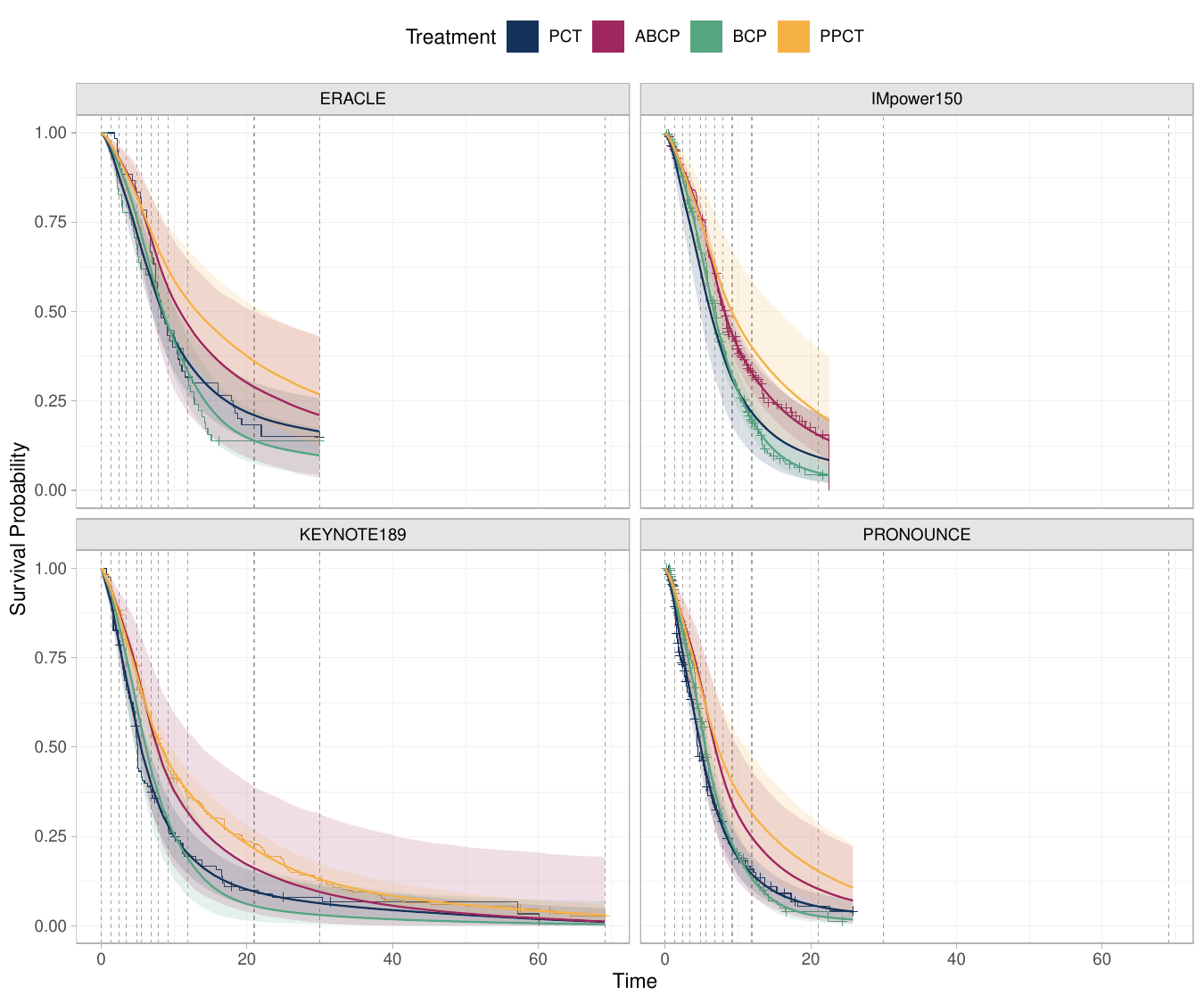}
  \caption{Estimated progression-free survival curves on each treatment in each study population, from a cubic M-spline model with an increased number of knots (11 internal knots). Non-proportional hazards are modelled via treatment effects on the spline coefficients. Shaded bands indicate 95\% credible intervals. Observed Kaplan-Meier curves are also shown. Dashed vertical lines indicate the location of knots.}
  \label{fig:nsclc_survival_nph_11kt}
\end{figure}

\begin{figure}
	\centering
  \includegraphics[width=\textwidth]{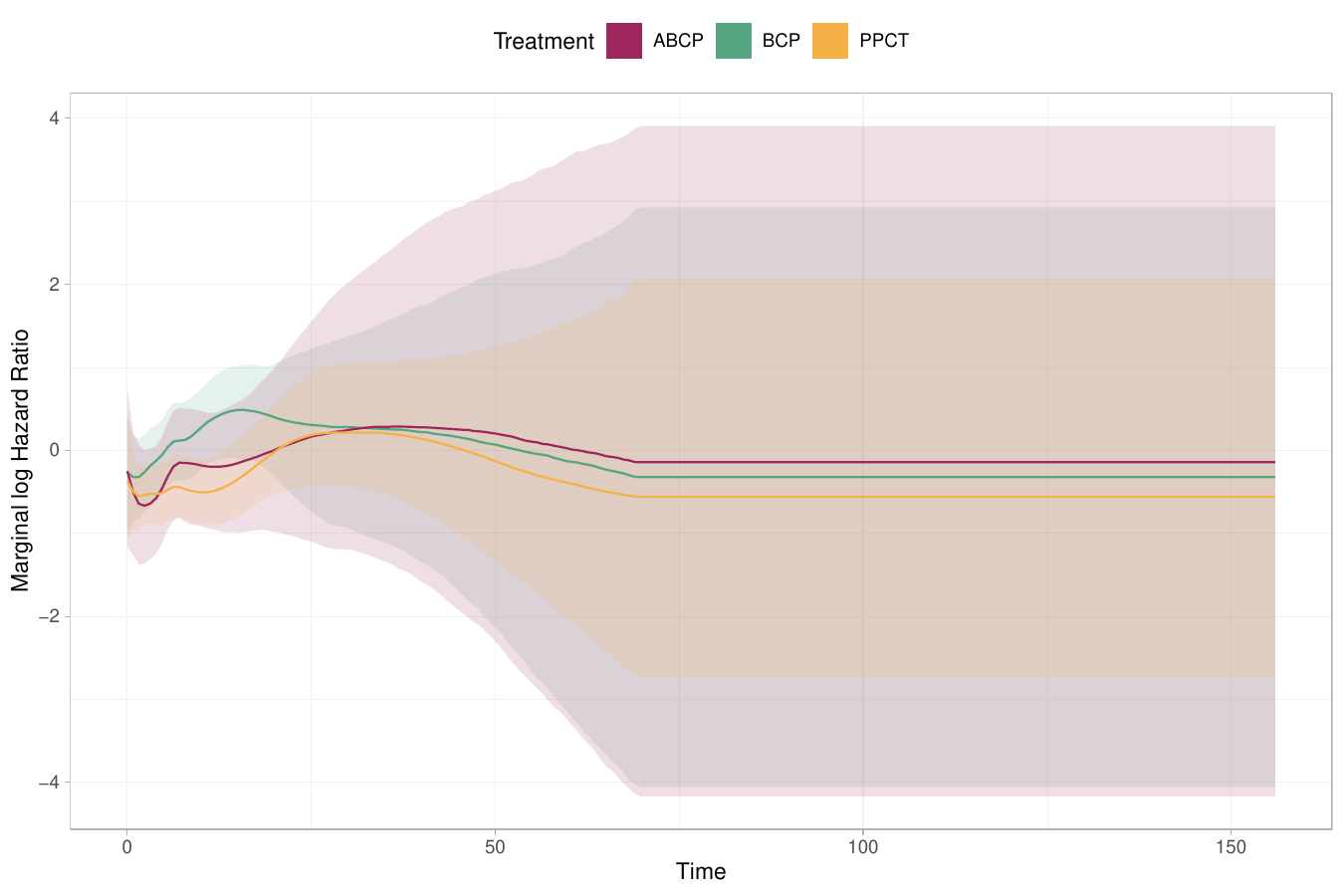}
  \caption{Estimated marginal log hazard ratios on each treatment over time in the KEYNOTE189 population, from a cubic M-spline model with non-proportional hazards via treatment effects on the spline coefficients. Shaded bands indicate 95\% credible intervals.}
  \label{fig:nsclc_keynote_hr}
\end{figure}

\end{document}